\documentclass[runningheads]{llncs}
\usepackage[title]{appendix}
\usepackage{graphicx}
\usepackage[dvipsnames]{xcolor}
\usepackage{tabularx,colortbl}
\usepackage{tabularray}
\usepackage{afterpage}

\usepackage{smartdiagram}
\usesmartdiagramlibrary{additions}

\usepackage{tikz}

\usepackage{pgfplots}
\pgfplotsset{compat=1.18}
\usepgfplotslibrary{statistics}

\usepackage{subcaption}
\usepackage{caption}

\usepackage{tabularx}

\usepackage{xcolor}
\definecolor{ou-red}{cmyk}{0,0.89,0.85,0.11}
\definecolor{ou-red-darker}{cmyk}{0,1,0.96,0.29}

\usepackage{amsmath}
\usepackage{amssymb}
\usepackage{csquotes}
\usepackage{mathtools}

\usepackage{hyperref}

\usepackage{caption}
\usepackage{multirow}
\usepackage{tcolorbox}
\tcbuselibrary{skins,breakable}

\definecolor{thesis-main}{HTML}{444444}
\definecolor{thesis-main-light}{HTML}{ECECEC}
\definecolor{thesis-blue-light}{HTML}{D8ECFF}
\definecolor{thesis-blue-dark}{HTML}{3A6FD4}
\definecolor{thesis-purple-light}{HTML}{EBDDFB}
\definecolor{thesis-purple-dark}{HTML}{7B2D8B}
\colorlet{main}{thesis-main}
\colorlet{main-light}{thesis-main-light}
\colorlet{blue-dark}{thesis-blue-dark}
\colorlet{blue-light}{thesis-blue-light}
\colorlet{purple-light}{thesis-purple-light}
\colorlet{purple-dark}{thesis-purple-dark}

\newtcolorbox{custombox}[1][]{%
  enhanced,
  frame hidden,
  boxrule=0.8pt,
  colframe=main!70!black,
  interior style={color=main-light},
  boxsep=0pt,
  left=5mm,right=5mm,top=3mm,bottom=3mm,
  attach boxed title to top left={yshift=-2mm,xshift=2mm},
  boxed title style={
    enhanced,
    boxrule=0.8pt,
    colframe=blue-dark,
    interior style={color=main},
    fontupper=\bfseries\color{white},
    left=3mm,right=3mm,top=1mm,bottom=1mm
  },
  overlay={%
    \begin{scope}[on background layer]
      \shade[rounded corners=4pt, left color=purple, right color=blue]
        (frame.south west) rectangle (frame.north east);
      \shade[rounded corners=5pt, left color=purple-light, right color=blue-light]
        ($(frame.south west)+(1.4pt,1.4pt)$) rectangle ($(frame.north east)-(.8pt,.8pt)$);
    \end{scope}
  },
  before upper={\begingroup\setlength{\leftmargini}{1.5em}},
  after upper={\endgroup},
  width=\textwidth,
  #1
}

\usepackage[para]{footmisc}

\usepackage{enumitem}    
\newlist{compactenum}{enumerate}{4}
\setlist[compactenum,1]{nolistsep}

\usepackage{booktabs}

\usepackage{placeins}

\newcommand{\citep}[1]{\cite{#1}}
\newcommand{\citet}[1]{\cite{#1}}

\AtBeginDocument{
\addtolength{\abovedisplayskip}{-1ex}
\addtolength{\abovedisplayshortskip}{-1ex}
\addtolength{\belowdisplayskip}{-1ex}
\addtolength{\belowdisplayshortskip}{-1ex}
}

\begin{document}

\title{Capturing Design Intent Before The Code}%\thanks{Supported by organization x.}}

 \author{Jens Krijgsman\inst{1}\orcidID{0009-0002-0603-7261} \and \\ Gideon Maillette de Buy Wenniger \inst{1,2}\orcidID{0000-0001-8427-7055} \and \\ 
 Clara Maathuis \inst{1}\orcidID{0000-0003-3483-1569}
% }
}

\authorrunning{J. Krijgsman et al.}
\institute{Open Universiteit, Heerlen, The Netherlands \\
\email{jens.krijgsman AT howest.be}\\\email{gemdbw AT gmail.com}\\
\email{clara.maathuis AT ou.nl}\\
\and
University of Groningen, Groningen, The Netherlands \\ 
}
%
% https://tex.stackexchange.com/questions/299813/removing-and-in-the-author-list
\renewcommand{\lastandname}{\unskip,}

\maketitle              % typeset the header of the contribution

\begin{abstract}
Generative AI is transforming software development, but its impact on early stage engineering practices remains limited. This contributes to a growing disconnect between architectural intent and implementation. To this end, this research introduces Sdlicit as a knowledge-grounded multi-agent framework designed to enforce traceability across software engineering artifacts throughout the design phase. Integrated with Visual Studio Code and exposed through the Model Context Protocol (MCP), the system supports the transformation of unstructured client briefs into structured Statements of Work, Software Requirements Specifications, architecture decisions, and parser valid Gherkin scenarios. The evaluation shows that LightRAG-based grounding over engineering standards achieves high retrieval reliability, yielding a RAGAS faithfulness score of 0.974. Moreover, by maintaining artifact relationships by construction, Sdlicit produces dependency networks 5.4 times denser than conventional workflows (85.6 versus 15.8 links). Additionally, replay comparison of generated architectural decision records based on an input brief showed a fair amount of matching against ground-truth values;  as assessed by a LLM-as-a-judge based evaluation. 
These findings reflect that knowledge-grounded and traceability-centric AI systems have the potential to strengthen pre-implementation software engineering processes and mitigate documentation drift.


\end{abstract}

\section{Introduction}
\label{ch:introduction}
The discrepancy between short-term productivity gains and long-term maintenance degradation shows a vulnerability in contemporary AI-assisted software engineering workflows. This tension can be partly explained by differences in task context. In greenfield settings, where software artifacts are generated under relatively few architectural and historical constraints, AI systems can accelerate the production of isolated components. Consistent with this observation, previous studies report productivity improvements of 55.8\% for tasks involving new software artifacts, particularly in early-stage or greenfield development \cite{Peng2023}.

Nevertheless, the industrial software development is predominantly conducted within brownfield environments characterized by legacy dependencies, accumulated design decisions, implicit architectural trade-offs, and evolving developer mental models. This architectural exposure contributes to progressive information loss across the Software Development Life Cycle (SDLC) \cite{Becker2025,dora2025state}. In addition, the design rationale and requirements elicited during early stakeholder interactions often fail to persist as structured development artifacts and are dispersed across informal conversations, meetings, chat logs, or undocumented decisions. Hence, when developers or autonomous coding agents later attempt to maintain, extend, or refactor these systems, they must expend substantial cognitive effort reconstructing the original intent that should have been preserved throughout the development process.
A similar problem arises during project handover when new personnel needs to infer the system’s logic from partial documentation and dispersed implementation artifacts. 
As such, in brownfield environments, both Generative AI systems and human implementers are prone to errors and failure when they lack sufficient access to the historical and contextual knowledge that governs the codebase. 
Similarly, when generative AI tools are applied to these environments, which help with code generation and related tasks, their impact could become negative, e.g., with reported productivity reductions of up to 19\% \cite{Becker2025}.
This productivity paradox can be attributed to the limited ability of current AI systems to include repository wide context and historical design rationale. Addressing this limitation requires designing methods that capture, preserve, and propagate the system’s mental model throughout the SDLC without relying on model fine-tuning \cite{Becker2025,dora2025state}. Along these lines, existing AI-based software engineering (SE) tools can only offer limited help with code-generation and related tasks in brownfield environments, because they struggle to maintain coherent representations of dependencies and changes across multiple files over time \cite{Gautam2025}. A way to tackle this is by providing direct access to internal organizational data and contextual information to the AI tools developed \cite{dora2025state}.

To address this limitation, this research aims to capture, structure, and preserve user intent during the early SDLC phases of planning, requirements elicitation, and architectural design to enable subsequent development and maintenance activities so that a reliable and verifiable record of that intent can be contained. Accordingly, this research proposes the Sdlicit framework (whose name combines SDLC and solicit) to actively elicit and refine user reasoning about software development decisions.\footnote{The Sdlicit source-code is available from: \url{https://github.com/JensKrijgsman/Sdlicit}}
The framework integrates several complementary techniques such as Socratic prompting \cite{Chang2023}, Chain-of-Thought (CoT) reasoning \cite{Wei2022},  Theory of Mind (ToM) \cite{Zhou2025}, Retrieval-Augmented Generation (RAG) \cite{Lewis2020}, and the Model Context Protocol (MCP). Sdlicit is implemented as an RAG system supported by a curated knowledge base of design patterns and IEEE standards. It uses Socratic prompting to guide software engineers in producing Software Requirements Specifications (SRS) and Architecture Decision Records (ADR), and subsequently derives Behavior-Driven Development specifications (BDD) in Gherkin syntax. By incorporating the BDD scenario generation \cite{North2006,Solis2011}, the framework strengthens the connection between requirements, architectural rationale, and executable development guidance. Furthermore, the system considers explainability as a design requirement as each generated recommendation is linked to an identifiable knowledge base source and citation faithfulness is assessed where access to model internal information permits verification.

The underlying research question formulated in this research is: How can an AI system assist software engineers in creating high quality architectural artifacts and how can the effectiveness of this system be evaluated? To address it, a Design Science Research methodological approach \cite{vomBrocke2020} is followed in which the proposed artifact is evaluated through qualitative case studies and prompt evaluation frameworks. A central design principle of Sdlicit is that AI-generated suggestions need to be both plausible and traceable in terms of system components and input source documents responsible for generation;
so that users can inspect, evaluate, and contest the outcome provided. This is important for the development of trustworthy AI-assisted SE systems \cite{Wallat2024}. 

The contributions of this research are further defined. First, it introduces the Sdlicit framework architecture that repositions AI from a passive autocomplete mechanism to an active Socratic collaborator by supporting the AI and SE communities in their effort on transforming high-level software requirements specifications into structured ADR and executable BDD specifications while preserving elicitation structure and traceability. 
And second, it provides an empirical evaluation protocol based on simulated repository environments and architectural case studies through a combination of deterministic syntax checks, statistical text-similarity metrics, and LLM-as-a-judge assessments to create a reproducible approach for evaluating artifact generation systems.

The remainder of this paper is structured as follows. Section 2 discusses the background and related studies in this domain. Section 3 presents the design of the system proposed. Section 4 presents the various experiments including the used evaluation procedures and metrics. Section 5 provides concluding remarks and discusses future research perspectives. 

\section{Related Work}
\label{ch:relatedwork}
In this section we discuss the most relevant related work for intent-driven software development, across three domains. We refer to Appendix \ref{appendix:additional-backround} for an extended discussion, which also covers the use of AI in software engineering.

\subsection{The evolution of the Software Development Life Cycle (SDLC)}
The waterfall method for the SDLC \citep{Royce2021}  proposes an iterative process for software development consisting of coding, testing, and operation steps with a clear focus on through documentation. In contrast, methodologies such as the Agile method \citep{Cockburn2001} were created to improve the management of software development, shifting the focus towards processes, tools, and working software and away from comprehensive documentation. Additionally, various variants (traditional, Agile, or hybrid) of SDLC processes were proposed \citep{Shah2016}.
These methodologies share several characteristics that are relevant to this research:
\begin{compactenum}[label=(\alph*)]    
    \item Using five phases: Requirements, Design, Coding, Testing, and Deployment.
    \item Suffering from communication gaps between stakeholders and development teams.
    \item Struggling with un-managed evolving requirements where changes occur but tracking them proves difficult.
    \item Benefiting from customer feedback during application.
\end{compactenum}

Test-Driven Development (TDD) \citep{Beck1999} adds another perspective which
follows a red-green-refactor cycle where developers write failing tests before implementation. Writing tests before code ensures that requirements are explicitly captured in executable form, which directly relates to the artifact-driven approach proposed in this work. Behavior-Driven Development (BDD) was introduced as a refinement of TDD that emphasizes collaboration between developers, testers, and business stakeholders \citep{North2006}. This uses a domain-specific language called Gherkin, which employs natural language constructs to describe system behavior in a format understandable by both technical and non-technical team members \citep{Solis2011}. Gherkin is chosen over alternative specification formats such as plain markdown, UML sequence diagrams and HTML-based systems because it satisfies three requirements of this research simultaneously: be human-readable, be machine-parseable to allow automatic validation, and having mature tooling. The corresponding tool (Cucumber\footnote{\url{https://cucumber.io/}}, Behave\footnote{\url{https://behave.readthedocs.io/en/latest/}}) enables direct execution as tests which closes the gap between documented intent and verified implementation \citep{wynne2017cucumber}. 

\subsubsection{Structured Artifacts in Software Engineering}
The practice of documenting architectural decisions has been formalized through Architecture Decision Records \citep{nygard2011}. ADRs capture the context, decision, and consequences of significant architectural choices, creating a traceable record of design rationale.
This documentation practice addresses what Tyree and Akerman identify as a critical gap: the loss of architectural knowledge over time \citep{Tyree2005}.
Multiple ADR templates exist, including the original Nygard format (Context, Decision, Status, Consequences), the Markdown Any Decision Records (MADR) format which adds structured options analysis, and the Y-Statement format for concise decision capture \citep{Kopp2018}.
The choice of template affects both the ease of creation and the completeness of captured rationale.
Software Requirements Specifications (SRS) provide another form of structured artifact, standardized by IEEE 830-1998 \citep{ieee830} and its successor ISO/IEC/IEEE 29148:2018 \citep{ieee29148}. This defines requirements for unambiguity, completeness, consistency, verifiability, and traceability. In these lines, recent research investigated which quality defects in requirements artifacts are automatically detectable, finding that completeness, consistency, and unambiguity can be partially automated through requirements smell detection \citep{Femmer2017}.
This provides a foundation for evaluating AI-assisted requirements artifacts.

\subsection{LLMs, RAG, LLM Agents, Reasoning and Acting}
Large language models (LLMs) are increasingly used in the software engineering domain. Here, Retrieval-Augmented Generation (RAG) \cite{RAG_NEURIPS2020} addresses a fundamental LLMs limitation: their knowledge is frozen at training time so they can generate plausible-sounding but factually incorrect information, i.e., hallucinations.
By first retrieving the most relevant document chunks from a database, and adding these as additional context to the input, RAG systems can reduce hallucinations. However, while traditional RAG excels at pinpointing isolated text snippets, its flat retrieval of independent text chunks loses the global structure and connected document dependencies characteristic of software architectures. 

To capture these relationships, GraphRAG converts raw text into an interconnected graph of entities, attributes, and relationships \citep{Edge2024}. Systematic comparisons show that GraphRAG benefits queries that depend on global, inter-dependent context, whereas plain RAG exhibits complementary strengths across task types \citep{Han2025}. Nonetheless, traditional GraphRAG has a high computational cost, uses a large amount of LLM calls, and furthermore focuses only on global queries, overlooking local queries which  traditional RAG covers effectively. 
To overcome this limitation, the Sdlicit framework incorporates the dual-level retrieval paradigm established by LightRAG, covering both local and global queries while increasing efficiency 
\footnote{\url{https://github.com/HKUDS/LightRAG}} \citep{Guo2025}. 

\subsubsection{Reasoning and Acting in LLM Agents} 
The foundational technique of Chain-of-Thought (CoT) prompting enables models to tackle complex tasks by generating a series of intermediate reasoning steps instead of directly predicting a final answer \citep{Wei2022}.
Building on this, sequential decomposition methods enhance reasoning by breaking complex problems into manageable sub-questions.
Recent advancements like Massively Decomposed Agentic Processes (MDAPs) show that extreme decomposition coupled with multi-agent voting and error correction at each step, can solve long-range tasks with higher reliability \citep{Meyerson2025}.
Such decomposition is vital for architectural trade-offs, where broad design goals must be partitioned into specific components to avoid the accumulation of errors.

\subsubsection{Synthesizing the Knowledge Gap}
Reviewing the literature reveals a major missing link at the intersection of architectural documentation, cognitive prompting architectures, and verified grounding. While frameworks like Elicitron and RECOVER \citep{Voria2025} automate upfront requirements gathering, they treat these insights as isolated assets which fails to propagate them downstream to guide implementation tools \citep{Mohammadmehdi2024,Voria2025}. At the same time, code generation models evaluated via benchmarks like SWE-bench operate almost entirely at the lower syntax layer, missing the high-level design rationale captured in documents like ADRs \citep{Jimenez2024}. Furthermore, standard flat vector methods fail to preserve global architectural dependencies, while traditional graph-based retrieval introduces high performance costs that make interactive development impractical. 

To address these points, the Sdlicit framework proposes a multi-stage architecture with the following characteristics. First, to counter artifact isolation and the low-level focus of coding tools, the pipeline establishes a continuous lineage that enforces traceability from initial client requirements down to executable behavioral specifications. Second, the global design dependencies missed by flat vector databases are captured via LightRAG's dual-level retrieval graph, removing the high computational costs of traditional graph traversal. 
Consequently, Sdlicit provides a cohesive pipeline that protects design intent across the entire software development lifecycle.

\begin{figure}[t]
\centering
\includegraphics[width=0.75\textwidth,keepaspectratio]{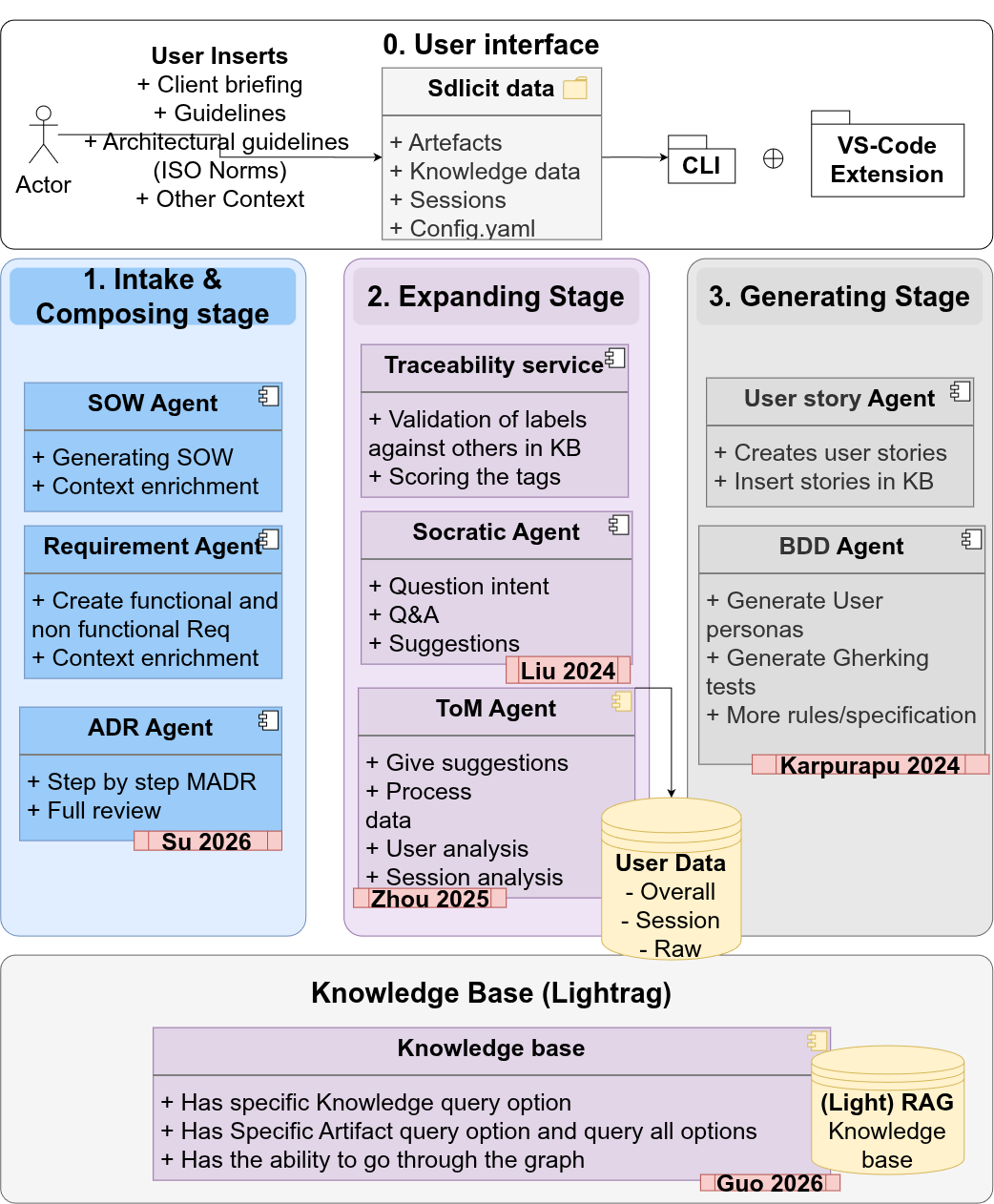}
\caption{Sdlicit architecture: a three-stage pipeline (Composing, Expansion, Generation) on top of a central LLM Gateway, with cross-cutting Socratic and Theory-of-Mind services and a dual-store Knowledge Base. The orchestrator is the single composition root.}
\label{fig:GeneralArchitecture}
\end{figure}

\section{Methods}
The methodological choices made in this research can be understood from the requirement of creating a traceable architectural artefact generation (AAG) system following the design science methodology. Whereas classic data science heavily relies on availability of labeled data for model training and testing; parts of the AAG task involve human interaction, making automatic evaluation inapplicable to those parts. At the same time, over-reliance on human evaluation is not practical for the evaluation of the developed system; since it involves  many hyperparameter choices at every stage, driving up the cost in terms of time and money of human evaluation. As such, automatic evaluation is essential, and part of the contribution of this work is also in proposing scalable ways to automatically evaluate various aspects and components of the system for which this is possible. Importantly, the push for automatic evaluation also coincides with the requirement of traceability, and aligns well with standard software design requirements regarding software modularity.
With these considerations and requirements in mind, the system architecture and implementation discussed next lays the foundation for enabling the effective (automatic) evaluation discussed in subsection \ref{sec:eval-sq2a}.

\vspace{-0.5cm}
\subsection{SQ1: The Sdlicit System Architecture}
\label{sec:system-architecture}

Sdlicit is a human-in-the-loop architectural assistant. Its guiding principle is that AI that supports software engineering should not be allowed to block architectural thinking by software engineers.
For this reason, architectural decisions should be made by the software engineers; or at least in full consultation of and confirmation by them. Therefore, before any implementation, an artefact should be produced and the engineer is required to make their \emph{intent} explicit and to defend it. 

The system is built as a decoupled client-server application. 
The backend's logic is partitioned into three sequential stages, each a separate package in the source tree:
\begin{enumerate}[noitemsep,topsep=0pt,parsep=0pt,partopsep=0pt]
    \item \textbf{Composing} (\texttt{composing\_stage/}) converts an unstructured client brief into a formal Statement of Work (SOW), then guides the engineer field-by-field through foundational Architecture Decision Records (ADRs).
    \item \textbf{Expansion} (\texttt{expansion\_stage/}) a cross-cutting phase where the Socratic Solicitor challenges assumptions, the Theory-of-Mind service calibrates the system's pedagogical depth, and the Traceability Service analyses the growing artefact graph.
    \item \textbf{Generation} (\texttt{generation\_stage/}) translates the approved high-level intent into deterministic downstream artefacts: a Software Requirements Specification (SRS), personas, user stories, and Behavior-Driven Development (BDD) scenarios.
    In-loop validation: \texttt{dspy.Refine} loops are used to automatically improve the output of the generation components by retrying, when deemed necessary based on output-validation by the system. 
    
\end{enumerate}

For the remainder of this work these functional modules are referred to as ``agents''. They do not run fully autonomous loops like multi-agent learning systems, but they operate via specific LLM requests augmented with tool access and memory retrieval. The term aligns with current literature on LLM-based software-engineering systems~\cite{Li2025} and avoids both overstating their autonomy and understating their statefulness.
 
\subsubsection{Component Structure}
\label{subsec:sys-components}
 
The Sdlicit architecture is implemented a system that consists of a three-stage pipeline, shown in Figure~\ref{fig:GeneralArchitecture}. A single orchestrator composes a small number of services and agents, every external model call passes through one gateway, and every artefact and memory file lives on disk under \texttt{<project>/.sdlicit/}. 
This structure keeps the system \emph{horizontally composable}. Adding a new artefact type requires only four local changes: registering a chunker, declaring a DSPy signature, optionally registering a validated program, and wiring the agent into the orchestrator's construction order.
No other changes would be needed. The same property is what makes the architecture testable: because the orchestrator injects every dependency.

Whereas due to space restrictions the main text of the paper cannot do full justice to the system architecture, and the architectural decisions and rationale; more details are provided in Appendix \ref{appendix:system-architecture}.

\subsubsection{Retrieval-Augmented Generation}

The system uses LightRAG, with a knowledge base storing both static documents and documents produced from the current session. The static reference material consists of international standards (ISO) documents, which establish industry best practices for engineering, systems development, usability, and security.\footnote{Specficially, the ISO/IEC/IEEE 29148 [IEEE, 2018], ISO 9241-210 [ISO,2019], 
ISO/IEC/IEEE 12207 [ISO, 2017], ISO/IEC/IEEE 42010 [IEEE, 2011] and ISO/IEC 27001 [iso, 2013]) documents were used.} 
The session-produced documents include the statement-of-work, ADRs, personas, and stories). Both document stores share a single LightRAG graph, which is superior to two separate stores, both in terms of computational efficiency and retrieval effectiveness.

\FloatBarrier

\section{Experiments}
\label{sec:experiments}

\subsection{Validation of retrieval-augmented generation (RAG) approach}

In a pre-study to the rest of the research, we used the RAGAS framework~\citep{Es2024} to validate the value of using an external knowledgebase in combination with retrieval augmented generation (RAG) as part of the SDLICIT System. The detailed results of this study are discussed in Appendix \ref{appendix:knowledge_retrieval_and_grounding_quality}. RAGAS results suggest that RAG manages to retrieve the relevant context in this domain; with for the graph-RAG retrieval system RAGAS scores of 0.974 (Faithfullness),  0.754 (Answer Relevancy), 0.744 (Context Recall) and 0.773 (Context Precision). A limitation is that RAGAS metrics by design cannot show attribution of the generation to the retrieved context.

\FloatBarrier

\subsection{Automated Architectural Artefact Quality Assessment without available Ground-Truth}
\label{sec:eval-sq2a}

The aim of the experiments discussed in this subsection is to test the Sdlicit system on a new greenfield software development scenario, and judge its performance as well as possible in the realistic situation of not having a ground-truth for the outputs generated by the system.
To do this end-to-end system testing, Sdlicit is used to design a time-tracking application integrated with GitHub and Outlook (client brief
in Appendix~\ref{sec:appendix-timetrack-brief}). The input for the test is a \emph{Scenario Brief}, corresponding to the client interview summary, providing the purpose, context and informally formulated requirements for this application.  
This time-tracking scenario feeds the pipeline-vs-baseline comparison. For the evaluation phase, a clean session is run from scratch to produce a full artifact set (SOW, requirements, ADRs for database selection, user stories, BDD scenarios).

Lacking a ground-truth for outputs, the Sdlicit system can still be evaluated in various ways. First it is evaluated in terms of 
\emph{structural trace coverage}, a concept measuring the syntactic coherence of generated named outputs that are cross-referenced. In addition, the apparent \emph{semantic (trace) coverage} of generated requirements by later generated artefacts is evaluated.    
Finally, link density metrics are inspected as well as the syntactic correctness of generated BDD scenarios. Jointly, these metrics provide a multi-faceted view on system performance subject to the requirements of being automatically computable and not requiring a ground-truth output for comparison. These evaluation metrics are next described in more detail.

\subsubsection{Metrics}
\label{sec:eval-sq2a-metrics}

\paragraph{Structural trace coverage} evaluates the mechanical cross-referencing chain
between stages of the architectural design process (requirements, ADRs, user stories, BDD
specifications). Artifact identifiers (e.g. REQ-001, ADR-02) are extracted from the generated
documents using regular expressions. For each ID it is then checked if the ID  exists in the workspace, meaning that it is already defined somewhere, either as system input (e.g. requirements) or intermediate result in a prior generated document. The structural traceability score $T_{\text{struct}}$ is defined as: 
$$T_{\text{struct}} = |E_{\text{valid}}| / |E_{\text{total}}|$$
where $E_{\text{total}}$ is the set of back-references the workspace attempted (for example, an ADR saying \texttt{implements: [REQ-AUTH-01]}), and $E_{\text{valid}}$ is the subset whose targets actually exist somewhere in the workspace.

\paragraph{Link density metrics} A workspace with fifty back-references, all valid, gets $100\%$ score for $T_{\text{struct}}$ , but so does a workspace with just two back-references that are both valid. 
To tell these apart the following is also reported:
\begin{itemize}
    \item \texttt{total\_links}, \texttt{valid\_links}, \texttt{broken\_links}: count of attempted back-references and counts of the corresponding subsets of links that reference existing vs. non-existing artifacts.
    \item \texttt{n\_req\_ids}: count of distinct REQ IDs that appear anywhere in the workspace.
    \item \texttt{n\_kb\_citations}: count of distinct \texttt{KB:topic:source} markers.
\end{itemize}

To get a better idea of the true performance, the link density metrics must be considered together with the structural trace coverage metric.  A classic information-retrieval recall score is not available here, because the true denominator of necessary links is unknowable: different engineering teams can legitimately design entirely different architectural topologies from the same brief.

\paragraph{Gherkin Validity}
The Sdlicit model first generates Decision Records and Software Requirements Specifications; it then generates Behavior-Driven Development (BDS) scenarios in Gherkin format from those artifacts. 
Whereas without any ground-truth for these outputs, a ``deep'' evaluation of these generated is not feasible, a superficial validity check still is. Specifically, \emph{Gherking Validity} is determined, by checking if the produced  files parse under the official Cucumber Gherkin parser. We additionally report the \emph{gherkin\_total}: the number of features that appear in the  
gherking BDS scenarios.

\paragraph{Semantic trace coverage} 

While structural traceability ensures that explicit identifiers are maintained, semantic traceability evaluates whether the core concepts of those requirements are preserved within the downstream artifacts. A composite text similarity score $S_{\text{sem}}(d_{\text{src}}, d_{\text{tgt}}) \in [0,1]$ is computed between a source requirement $d_{\text{src}}$ and a generated downstream artifact $d_{\text{tgt}}$:

\vspace{-0.5cm}
\begin{align}
& S_{\text{sem}}(d_{\text{src}}, d_{\text{tgt}})
=  \nonumber \\
& \alpha \cdot \text{sim}_{\text{tfidf}}(d_{\text{src}}, d_{\text{tgt}})
\ + \ \beta \cdot \text{sim}_{\text{jaccard}}(d_{\text{src}}, d_{\text{tgt}})
\ + \ \gamma \cdot \text{sim}_{\text{topic}}(d_{\text{src}}, d_{\text{tgt}})
\end{align}

\noindent The three weighted features measure different things:
\begin{itemize}
    \item \textbf{TF-IDF cosine.} TF-IDF stands for term frequency, inverse document frequency. Each word in a document is weighted by how often it appears in that document divided by how often it appears anywhere in the corpus. Rare technical words count more. Common stopwords count for nothing. Cosine similarity between two TF-IDF vectors measures shared rare vocabulary on a $[0, 1]$ scale.
    \item \textbf{Keyword Jaccard.} Treats both documents as bags of (stopword-filtered) words and asks $|A \cap B| / |A \cup B|$. Captures concept word overlap regardless of frequency.
    \item \textbf{NMF topic alignment.} NMF (non-negative matrix factorisation) is a topic modelling technique that 
    represents each document as a mixture of latent topics learned from the corpus. Cosine similarity in this topic space captures concept level alignment even when the surface vocabulary differs.
\end{itemize}

As feature weights we used $\alpha=0.5, \beta=0.3, \gamma =0.2$. Furthermore, a threshold $\tau = 0.3$ is used, and a requirement is considered semantically covered if at least one downstream artefact passes this threshold when compared against the requirement. 
The feature weighting is heuristic, based on observations: TF-IDF dominates because rare-word overlap is the most discriminative signal, NMF is smallest because its topic estimate is noisy on small workspaces. The threshold $\tau = 0.3$ is calibrated to sit above coincidence (about $0.1$) but below verbatim quoting (about $0.6$).

\subsubsection{Experimental Setup}

While an important part of the full system, in all the experiments in this paper the Socratic Agent is disabled, as it requires human user input and can therefore not be deployed as part of an automated evaluation. 
The pipeline can be ran under three conditions (full, lean, no-RAG). The full condition enables retrieval together with the advisory review layer (ADR review, requirement coverage check, and Theory-of-Mind assessment). The lean condition keeps retrieval but disables the advisory layer. The no-RAG condition disables retrieval entirely. The default system setup is the \emph{lean} condition. This condition is used in all experiments,  except for two Ablation Study experiment conditions (\emph{full}, \emph{no-RAG}) discussed below. 
For the experiments, the gpt-5.4-nano cloud model is used as the generator LLM. Each experimental setting is repeated over 5 runs, with metric scores averaged.

\begin{table}[t]
\caption{Pipeline vs.\ single-prompt baseline. Pipeline is pooled from two
independent $N{=}5$ batches (this comparison and the ablation study,
Table~\ref{tab:nb02_ablation}) for $N{=}10$. Baseline is $N{=}4$ (one of five
single-prompt runs failed schema validation).}
\label{tab:nb01_results}

\begin{subtable}[b]{1.0\textwidth}
\centering
\caption{Coverage metrics, gherkin validity and broken links.
``Significant?'' is at $\alpha = 0.05$ on the two-sided Mann-Whitney $U$
test. $r_{\text{rb}}$ is the rank-biserial effect size, on a scale from
$-1$ to $+1$.}
\label{tab:nb01_ensemble}
\resizebox{\linewidth}{!}{%
\begin{tabular}{lccccccc}
\toprule
\textbf{Metric} & \textbf{Pipeline} & \textbf{Baseline} & \textbf{$\Delta$} & \textbf{$U$} & \textbf{$p$} & \textbf{$r_{\text{rb}}$} & \textbf{Significant?} \\
 & $\mu$ \,($s$), $N{=}10$ & $\mu$ \,($s$), $N{=}4$ & & & & & \\
\midrule
Structural trace coverage (\%) & $100.0$ \,($0.0$) & $100.0$ \,($0.0$) & $0.0$   & $20.0$ & $1.000$ & $0.00$ & No \\
Semantic trace coverage (\%)   & $\phantom{0}54.3$ \,($22.0$) & $\phantom{0}40.6$ \,($31.5$) & $+13.7$ & $26.0$ & $0.454$ & $0.30$ & No \\
Gherkin validity (\%)          & $100.0$ \,($0.0$) & $100.0$ \,($0.0$) & $0.0$   & $20.0$ & $1.000$ & $0.00$ & No \\
Broken links                   & $\phantom{00}0.0$ \,($0.0$)   & $\phantom{00}0.0$ \,($0.0$)   & $0.0$   & $20.0$ & $1.000$ & $0.00$ & No \\
\bottomrule
\end{tabular}%
}
\end{subtable}
\begin{subtable}[b]{1.0\textwidth}
\centering
\caption{Link density supporting metrics. CV is the coefficient of
variation $s / \mu$.}
\label{tab:nb01_density}
\scalebox{0.9}{
\small
\begin{tabular}{lcc}
\toprule
\textbf{Metric} & \textbf{Pipeline} ($\mu \pm s$, CV), $N{=}10$ & \textbf{Baseline} ($\mu \pm s$, CV), $N{=}4$ \\
\midrule
\texttt{total\_links}          & $87.4 \pm 15.2$ \,(0.174) & $15.8 \pm 2.1$ \,(0.131) \\
\texttt{valid\_links}          & $87.4 \pm 15.2$ \,(0.174) & $15.8 \pm 2.1$ \,(0.131) \\ 
\texttt{broken\_links}         & $0.0 \pm 0.0$             & $0.0 \pm 0.0$             \\
\texttt{n\_req\_ids} (distinct)& $46.9 \pm 1.7$ \,(0.035)  & $\phantom{0}7.5 \pm 0.6$ \,(0.077) \\
\texttt{gherkin\_total} (features) & $\phantom{0}3.0 \pm 0.0$ & $\phantom{0}2.5 \pm 0.6$ \\
Wall-time (s)                  & $1938.0 \pm 473.9$        & $\phantom{00}68.0 \pm 40.8$ \\
\bottomrule
\end{tabular}
}
\end{subtable}
\end{table}

\paragraph{Single-Prompt Baseline Model}

The first part of the evaluation compares the Sdlicit pipeline against generation through a single-prompt baseline that does not use dspy.Refine and has no in-loop validation. In contrast to this baseline, the Sdlicit system has the ability to re-generate outputs if in-loop validation flags that the current outputs fail certain requirements.

\paragraph{Ablation study.}
The pipeline is run under all three conditions (\emph{full}, \emph{lean},
\emph{no-RAG}) to isolate the contribution of each subsystem that has a
measurable effect in a headless run (i.e. with no user input). 
The contrast between \emph{lean} and \emph{no-RAG} isolates the effect of retrieval. The contrast between \emph{full} and \emph{lean} isolates the advisory layer.

\subsubsection{Results}

\begin{table}[t]
\centering
\caption{Ablation results using the identical project brief and underlying model. \texttt{full} and \texttt{no-rag} are $N=5$. \texttt{lean} is pooled to $N=10$ (this study plus the pipeline-vs-baseline comparison, Table~\ref{tab:nb01_results}). 
}
\label{tab:nb02_ablation}
\small
\begin{tabular}{lccc}
\toprule
\textbf{Condition} & \texttt{full} ($N{=}5$) & \texttt{lean} ($N{=}10$) & \texttt{no-rag} ($N{=}5$) \\
 & $\mu$ ($s$) & $\mu$ ($s$) & $\mu$ ($s$) \\
\midrule
Structural trace coverage (\%) & $98.5$ ($2.0$) & $100.0$ ($0.0$) & $99.6$ ($0.5$) \\
Semantic trace coverage (\%)   & $67.3$ ($15.8$) & $54.3$ ($22.0$) & $58.0$ ($18.9$ \\
Gherkin validity (\%)          & $100.0$ ($0.0$) & $100.0$ ($0.0$) & $100.0$ ($0.0$) \\
Broken links                   & $1.2$ ($1.8$) & $0.0$ ($0.0$) & $0.4$ ($0.5$) \\
\bottomrule
\end{tabular}
\end{table}

\paragraph{Pipeline versus Baseline Comparison}

Looking at the results in Table \ref{tab:nb01_results}, both systems achieve $100\%$ trace coverage on every run. However, the paths are not equivalent. The pipeline achieves it by construction: its \texttt{dspy.Refine} loops reject any ADR that cites a non-existent requirement and retry with feedback until the citation is valid or the retry budget expires. The baseline achieves it by post-filtering broken links after generation. Without filtering invalid links, the baseline performance would collapse to roughly 50\% trace coverage. 

In terms of link density, the pipeline emits about $86$ total back-references per workspace. The baseline emits about $16$. The pipeline produces $5.4\times$ as many cross-references and $6.1\times$ as many distinct requirement IDs per workspace, and both at zero broken links.

Looking at semantic trace coverage, there is a substantial difference, with the pipeline performing slightly better (54.3 vs. 40.6), but it is not statistically significant. The standard deviation of this metric is also considerably lower for the pipeline than the baseline, however, both are really high indicating large differences between runs.
Gherkin validity is 100\% for both the pipeline and the baseline, indicating even single-prompt generation can generate syntactically correct gherkin BDD output using a sufficiently powerful LLM combined with effective prompt instructions.

\paragraph{Results Ablation Experiments}

Table \ref{tab:nb02_ablation} shows that  the advisory layer (\texttt{full} vs. \texttt{lean}) increases semantic coverage by $13.0$ points, however not statistically significant ($p = 0.310$). Retrieval (\texttt{lean} vs. \texttt{no-rag}) shows no benefit on this brief, with \texttt{no-rag} marginally higher on semantic coverage, again not a statistically significant difference ($p = 1.0$).

\FloatBarrier

\subsection{Replay Experiments: Assessing Architectural Artefact Quality with Ground-Truth}
\label{sec:eval-sq2b}

%%%% Setup
The automatic evaluation without ground-truth discussed in the previous section has many advantages. A strong aspect is that it could potentially also  be applied to let the system provide real-time user feedback for new scenarios. For testing the validity of the generated outputs it does however also have clear limitations.
To evaluate if Sdlicit is able 
to generate architectural artefacts that are in line with an established ground-truth, and furthermore see if 
it generalizes beyond the time-tracking case study, a replay study was performed on the \texttt{microsoft/semantic-kernel} repository. A replay study takes inputs from a real, existing project and compares the system's outputs against the project's actual decisions. Ten ADRs were manually selected from the repository, which contains a total of 59 ADRS, to span ten distinct architectural domains: API design, abstraction layering, project structure, AI integration, RAG, plugin design, connectors, storage, structured output, and JSON serialization. For each ADR the original problem brief was supplied as input, and the generated decision, alternatives, and citations were recorded. Two judges were applied: fuzzy Jaccard on alternatives (statistical text-similarity) and an LLM-as-a-judge categorical label (decision and alternatives quality).

\paragraph{Evaluation}
In the first part of the evaluation, the Brief of the reference ADR was provided to the system and a set of requirements (text lines) was generated. This set was compared against the same set from the reference ADR.

The \emph{fuzzy Jaccard index} was applied as the first evaluation axis to compare the alternative sets. Classical Jaccard requires exact element matching, which is too frail at the option-name level (``Use Maven'' and ``Adopt Maven as build tool'' should count as the same option). Two options are therefore treated as matching when their character-level similarity exceeds $\tau = 0.6$:
\begin{equation}
J_{\text{fuzzy}}(A, B)
= \frac{|\{(a, b) \in A \times B : \mathrm{sim}(a, b) \geq \tau\}|}{|A \cup B|}.
\end{equation}
The threshold admits paraphrase at the option-name level while rejecting
unrelated phrases. The Ratcliff-Obershelp ratio from \texttt{difflib.SequenceMatcher.ratio} is used to compute this. The comparison
is made over the \emph{whole} option-name strings, case-folded, with no
tokenisation and no stop-word filtering. In practice this remains a strict measure for this application.

\begin{table}[t]
\centering
\caption{Per record results from the Semantic Kernel replication study where the decision column indicates the automated agreement classification and the alternatives quality column details option space alignment. The \textbf{Jaccard} column is the thresholded fuzzy Jaccard ($\tau = 0.6$). The adjacent \textbf{Max sim} column reports the underlying \emph{closest} pairwise option similarity (the continuous Ratcliff--Obershelp ratio the threshold is applied to), which shows how near the option sets actually come. A worked example for ADR-0006 is given in Appendix~\ref{app:jaccard-worked}.}
\label{tab:sk_jaccard}
\resizebox{\linewidth}{!}{%
\begin{tabular}{lcccccll}
\toprule
\textbf{ADR} & \textbf{Topic (short)} & \textbf{GT \#} & \textbf{Gen \#} & \textbf{Jaccard} & \textbf{Max sim} & \textbf{Decision} & \textbf{Alts.\ Q.} \\
\midrule
ADR-0006 & Dynamic REST payload         & 2  & 3 & 0.00 & 0.27 & matches               & divergent \\
ADR-0008 & Generic LLM request settings & 3  & 4 & 0.00 & 0.38 & matches               & partial \\
ADR-0010 & .NET project structure       & 4  & 4 & 0.00 & 0.29 & different-justifiable & divergent \\
ADR-0017 & Function calling             & 3  & 4 & 0.00 & 0.42 & matches               & divergent \\
ADR-0034 & RAG in SK                    & 3  & 3 & 0.00 & 0.28 & different-justifiable & divergent \\
ADR-0039 & Plugin name metadata         & 3  & 4 & 0.00 & 0.36 & different-justifiable & divergent \\
ADR-0047 & OpenAI/Azure connectors      & 3  & 4 & 0.00 & 0.11 & matches               & divergent \\
ADR-0051 & EF vector store              & 2  & 4 & 0.00 & 0.35 & different-justifiable & partial \\
ADR-0053 & Structured outputs           & 11 & 4 & 0.00 & 0.35 & different-justifiable & divergent \\
ADR-0060 & JsonSerializerOptions        & 3  & 5 & 0.00 & 0.34 & different-justifiable & divergent \\
\midrule
\textbf{Aggregate} & & $3.7$ ($\pm 2.6$) & $3.9$ ($\pm 0.6$) & $0.00$ ($\pm 0.00$) & $0.32$ ($\pm 0.09$) & 4 m / 6 just.\ / 0 inc.\ & 8 div.\ / 2 partial \\
\bottomrule
\end{tabular}%
}
\end{table}

The second axis is an LLM judge (\texttt{openai/gpt-5.4} with high reasoning
effort). It judges both the produced decisions and alternatives, as generated based on the Brief. It emits two categorical labels per ADR: \emph{decision agreement}
(\texttt{matches} / \texttt{different-justifiable} /
\texttt{different-incoherent}) and \emph{alternatives quality}
(\texttt{full} / \texttt{partial} / \texttt{divergent}), each accompanied
by a free-text rationale. Two categorical labels are used rather than one
combined score because a composite hides which dimension results in a poor
result. The judge \texttt{gpt-5.4, medium thinking} is a larger and higher-reasoning model than the production
generator (\texttt{gpt-5.4-nano}), so it evaluates the artifacts from a
stronger vantage point than the system under test. The judge and the generator
belong to the same provider family; using a judge LLM from a different provider is left for future work.

\paragraph{Results}
Table \ref{tab:sk_jaccard} shows the results for the replay experiment. Note that 4 out of 10 of the decision sets are matching according to the LLM with 6 out of 10 falling in the `different-justifiable' category. In contrast, non of the compared decision records reach the fuzzy matching threshold, as shown by all the ``max sim'' scores being below 0.6; and consequently all the (fuzzy) Jaccard scores are zero for matching the decision records. The generated alternatives sets also differ from the reference according to the LLM judge, with partial overlap for only 2 out of 10 cases. This might be explained from the fact that given the Brief, the set of plausible of alternatives to draw from remains quite large, so the chance of the system choosing the same ones as in the reference is low. 
In contrast, the alternatives counts for the ground-truth (GT \#) vs. generated (Gen \#) are more similar, showing the system does manage to generate a conventional number of alternatives on average.

\FloatBarrier

\subsection{Multi-model experiments}

We ran additional experiments using different language models from different vendors, and with different model sizes. In these experiments we also tested with local models. 
The full results are provided in Appendix \ref{appendix:multimodel_experiments}.
In summary, we observed that the relatively larger cloud models (gpt-5.4-mini, gpt-4o) provided the best results. Much smaller models break the system, as expressed by lowered or even zero gherkin validity as well as structural trace coverage and semantic trace coverage. As such, the 4B parameters qwen3.5-4b model indicates the approximate minimal size of models to still give decent results. In comparison, the much larger (24B active parameters) mistral-small model improves structural and semantic trace coverage compared to this, but only achieves 60\% gherking validity; in line with the established wisdom that size alone is not the only factor determining model performance.

\section{Conclusions}
\label{ch:conclusions}

This paper introduces the Sdlicit framework to provide a concrete, multi-layered solution to the software engineering \textit{productivity paradox} by containing open-ended generative AI inside a structured, validated, and knowledge-grounded artifact pipeline. Implemented as a decoupled, human-in-the-loop client-server architecture, the system relies on declarative, typed DSPy signatures and Pydantic schemas wrapped in self-correcting refinement loops to validate engineering inputs before they reach disk. Grounding this pipeline with LightRAG dual vector-plus-graph retrieval raises RAGAS faithfulness from a baseline of 0.007 to 0.974, generating artifact dependency networks that are 5.4x denser than single prompt baselines. The empirical validation demonstrates that this architecture yields graph networks 5.4x denser than single-prompt baselines with 100\% trace coverage and complete Cucumber parser compliance when synthesizing Gherkin feature specifications from ADRs. These results show that embedding deterministic syntax filters, dual-mode retrieval, and multi-layered evaluation suites into the development lifecycle successfully halts documentation drift and eliminates broken references in the system design.

Conceptually, the research shows how structural and semantic trace coverage as well as link-statistics can be used to assess performance aspects without a system output ground-truth. Complementary, replay experiments can facilitate evaluation with output ground-truth.

As part of the Sdlicit research, a companion Visual Studio Code extension was created, facilitating actual use of the system. Future work will involve controlled user studies to measure the benefits of the proposed architecture.

Additionally, three more future research directions are considered. First, consider the continuation of the loop to implementation by developing backward-directed agent loops that parse git histories, pull request deltas, and abstract syntax tree (AST) commit diffs to reverse-engineer latent design rationale while simultaneously compiling generated Gherkin specifications into executable step definitions to calculate code-level structural trace coverage. Second, implement algorithmic prompt optimization and deployment sweeps by using deep meta-prompting algorithms like MIPROv2 or BootstrapFewShot to stabilize early stage software requirement synthesis, alongside exhaustive compute sweeps to map the performance floor across 7B to 14B parameter open weight models and their quantized variants. And third, considering the backend's MCP interface, a controlled benchmarking harness could be established to pit fully autonomous coding agents directly against the structured orchestrator in order to assess how unconstrained agentic generation impacts or degrades core architectural traceability and coherence over multi-session development lifecycles. 

\section*{Acknowledgments}
\subsection*{Declaration of Generative AI and AI-assisted technologies}
\credits{Large language models were used to assist with rephrasing and improving the clarity of this document. Tools like GitHub Copilot were also used in the creation of the Sdlicit system itself. AI has always been used in a way with the goal of strengthening the ideas and concepts, to understand decisions and shorten the `time to market'. For tasks like frontend development it is extensively used, more research-focused tasks were done in a more controlled step-by-step approach.}

\clearpage

\bibliographystyle{splncs04}
\bibliography{report}

% See: https://tex.stackexchange.com/questions/179997/using-package-appendix-with-llncs
\begin{subappendices}
\renewcommand{\thesection}{\Alph{section}}%

\setcounter{chapter}{2}
\setcounter{section}{0} % Reset to the first appendix section
\clearpage
\section{Additional Background}\label{appendix:additional-backround}

\subsection{LLMs, RAG, LLM Agents, Reasoning and Acting}

\subsubsection{Reasoning and Acting in LLM Agents}
A foundational framework for grounding LLM behavior in an external environment is Reasoning and Acting (ReAct), which combines reasoning with task-specific actions, enabling the model to interact with external environments such as APIs, databases, or command lines \citep{Yao2023}.
Dissimilar to static reasoning chains, ReAct operates through a dynamic ``Thought-Action-Observation'' cycle: the model generates a thought to plan or track subgoals, executes an action, and updates its context based on the external observation \citep{Yao2023}.
This grounding mechanism is critical for agentic software development, as it allows assistants to validate assumptions against the actual codebase, mitigate hallucinations common in pure Chain-of-Thought approaches, and recover from execution errors by refining the next thought \citep{Yao2023,Yang2024}.
In recent literature, new methods are suggested. For example, splitting up the task into many more subtasks can help solve more complex problems.
The MAKER framework, proposed by Meyerson et al., demonstrates that extreme decomposition into micro-agents allows systems to solve tasks requiring over a million steps with zero errors \citep{Meyerson2025}.
Other research emphasizes the need for stateful reasoning and structured cognitive architectures.
The Theory of Mind (ToM) framework for software engineering agents, proposed by Zhou et al.\ (2025), demonstrates that successful AI assistance requires maintaining a model of user intent, knowledge state, and goals throughout interactions \citep{Zhou2025}.
Their ToM-SWE approach shows that retrieval alone is insufficient without mechanisms to maintain long-term consistency with user intent, a finding directly relevant to the Sdlicit framework's goal of preserving architectural rationale across the SDLC.
Beyond retrieval, recent work has focused on programmatic prompt optimization.
Declarative Self-improving Python (DSPy)\footnote{\url{https://dspy.ai/}} treats prompts as modular functions that can be optimized against defined metrics, enabling systematic improvement of prompt quality without manual iteration \citep{Khattab2023}.
This approach is particularly relevant for building reliable AI systems where prompt consistency directly impacts output quality.

\subsubsection{Prompting Techniques for Reasoning}
Advanced prompting strategies have shifted the use of LLMs from simple text generation to complex reasoning, which is increasingly applied to support deliberate architectural decision-making.
These techniques aim to emulate ``System 2'' thinking, a slower and more methodical cognitive process that relies on logical analysis to solve intricate problems \citep{Zhang2025}.

\textbf{Socratic Inquiry and Requirement Elicitation.} \
Socratic inquiry represents a shift toward more proactive and recursive reasoning.
The ``Art of Socratic Questioning'' algorithm mimics human recursive thinking by leveraging a divide and conquer style, where the model proactively raises and addresses sub-questions until it gathers sufficient information to solve the original problem \citep{Qi2023}.
This method explicitly navigates the thinking space and is more robust than single-pass reasoning.
Complementary to this is the SocraticLM framework, which implements a ``Thought-Provoking'' paradigm to engage users in a dialogue rather than providing passive answers \citep{Liu-socratic2024}.
This approach is essential for requirements engineering, ensuring that stakeholders articulate implicit architectural constraints and resolve linguistic ambiguities before final decisions are made \citep{Bashir2025,Voria2025}.
The academic value of structured prompt design in software engineering is further supported by White et al., who catalogue a suite of reusable prompt patterns for tasks such as code quality improvement, refactoring, and system design, demonstrating that prompt structure has a measurable effect on the quality of LLM outputs \citep{White2023}.
This body of work underlines the Sdlicit framework's selection of Socratic and Chain-of-Thought patterns as the primary elicitation mechanism.

\textbf{Advanced Reasoning and Interactive Structures.} \
Beyond linear chains, robust planning frameworks allow models to explore multiple paths and verify their own logic.
Self-Consistency improves reasoning reliability by sampling multiple diverse reasoning paths and selecting the most consistent answer through majority voting \citep{Wang2022}.
These structures are often integrated into broader agentic paradigms that include planning, tool use, and feedback learning \citep{Li2025}.
For instance, the ReAct framework synergizes reasoning with external actions, enabling the model to retrieve real-world information and update its internal plans dynamically \citep{Yao2023}.
Furthermore, frameworks such as ``Reason for Future, Act for Now'' (RAFA) provide principled orchestration of reasoning and acting, using memory buffers to plan future trajectories while acting on current states \citep{Liu2023}.
This integration helps ground the design process, which reduces hallucinations and ensures that architectural decisions are supported by up-to-date evidence \citep{Raza2025}.

\subsection{Evaluation and Explainability}

Evaluating AI-assisted systems requires metrics that capture both output quality and the reasoning behind suggestions.
Explainability describes how well a system can make the reasoning behind its outputs understandable to the people who rely on them. In AI-assisted software engineering it is a prerequisite for responsible AI, because it underpins user trust, supports fairness, and helps meet regulatory compliance requirements \citep{Arrieta2019}.
Explainability is therefore essential for AI suggestions to be trusted, as users must be able to understand why recommendations were made before they act on them.
For retrieval quality, the Retrieval Augmented Generation Assessment (RAGAS)\footnote{\url{https://ragas.io/}} framework provides metrics including Context Precision for the relevance of retrieved documents, Context Recall for the completeness of retrieval, Faithfulness for the accuracy of generation relative to context, and Answer Relevancy \citep{Es2024}.
These metrics enable systematic evaluation of RAG systems.

While traditional, constraint-based prompt testing tools like Promptfoo\footnote{\url{https://www.promptfoo.dev/}} and DeepEval\footnote{\url{https://docs.confident-ai.com/}} provide valuable automated assertions for static prompt configurations, they were omitted in this work due to the adoption of Declarative Self-improving Python (DSPy)~\citep{Khattab2023}. Rather than manually adjusting raw prompt strings and writing unit tests to protect against output degradation, DSPy treats prompt engineering as a mathematical compilation pipeline. Prompts are parameterized as modular elements optimized against downstream performance metrics, rendering static regression testing frameworks redundant within this architecture. Consequently, the evaluation pipeline isolates prompt optimization under DSPy, while maintaining standard validation of retrieval quality via RAGAS.% and citation attribution via MIRAGE.

Traceability, which links each suggestion back to the source documents in the knowledge base, provides one form of explainability.
However, simply showing which documents were retrieved does not guarantee that the documents actually influenced the generation.
The presence of an explanation alone does not guarantee a correct decision either.
Research suggests that without mechanisms to force analytical engagement, users may still over-rely on incorrect AI outputs \citep{Bucinca2021}.
Determining whether an explanation is faithful to the model's internal process is therefore critical.
The Model Internals-based RAG Explanations framework (MIRAGE), introduced by Qi et al., addresses this citation faithfulness challenge by using model internals to provide faithful answer attribution \citep{Qi2024}.
MIRAGE detects context-sensitive answer tokens and pairs them with retrieved documents contributing to their prediction via saliency methods, achieving high agreement with human answer attribution.
The distinction between citation correctness, when the cited document supports the statement, and citation faithfulness, when the model actually used the cited document during generation, is crucial for trustworthy RAG systems \citep{Wallat2024}.
%%%%

\subsection{The Use of AI for Software Development}
\begin{figure}[!htbp]
\centering
\includegraphics[width=0.9\textwidth,keepaspectratio]{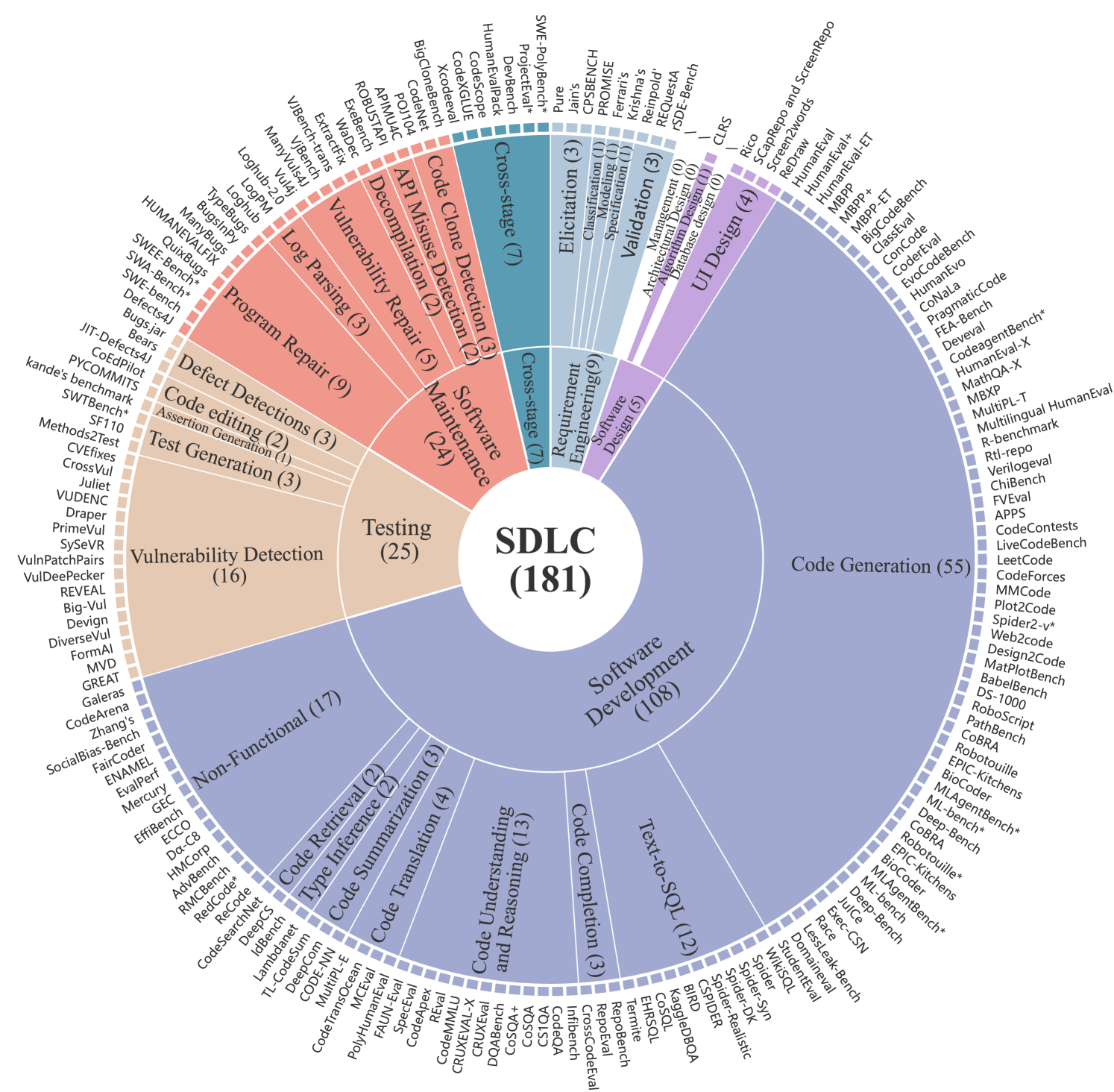}
\caption{Taxonomy of CodeLLM benchmarks across the SDLC.
This figure categorizes existing CodeLLM benchmarks according to different phases of SDLC.
The numbers indicate how many benchmarks correspond to each phase or task.
Entries marked with an asterisk (*) denote benchmarks that can be used to evaluate the capabilities of CodeLLM agents.
The blank spaces indicate that there are no relevant benchmarks for the current task.
This figure and the legend were obtained from \citep{Wang2025}}
\label{figWang}
\end{figure}

There is a high imbalance in research focus across the Software Development Life Cycle (SDLC).
From 181 benchmarks it was found that approximately 60\% focus exclusively on the software development (coding) phase, while other processes such as requirements engineering (5\%) and software design (3\%) remain under-represented (Figure~\ref{figWang}) \citep{Wang2025}.
This imbalance is also shown in the industry. The 2025 DORA report found that while 71\% of professionals use AI for writing new code, only 49\% utilize it for analyzing requirements.
This shows the gap in AI adoption \citep{dora2025state}.
Several frameworks have attempted to tackle these upstream phases of the SDLC. For instance, Elicitron is a simulation tool that generates diverse user agents to role-play product experiences, thereby uncovering latent requirements that traditional interviews might miss \citep{Mohammadmehdi2024}. Similarly, RECOVER utilizes natural language processing to automatically extract and classify structured system requirements directly from unstructured stakeholder conversations \citep{Voria2025}.
Both of these techniques are promising but do not propagate the generated requirements downstream to coding agents.
For this purpose, techniques such as the Model Context Protocol (MCP)\footnote{\url{https://modelcontextprotocol.io/}} could be used to create a multi-agent solution.
MCP provides a standardized interface for AI systems to access external data sources and tools, enabling integration with development environments.
This approach is better suited for context management and tool calling \citep{Jeong2025}.
For the later steps in the process, for example the refactoring step, there is also preceding research and benchmarks.
RefactorBench highlights that while agents can edit code, they significantly struggle with stateful reasoning, or the ability to track dependencies and changes across multiple files over time \citep{Gautam2025}.
Other relevant benchmarks, such as Galeras, focus on the causal interpretability of LLM performance, measuring how hidden variables, such as code complexity and prompt size, bias the model's ability to generate accurate code and summaries \citep{Rodriguez-Cardenas2023}.
The limitations highlighted by these benchmarks, the inability to track state (RefactorBench) and the susceptibility to confounding variables (Galeras), show the need for a structured context management system.
By ensuring that agents operate within a framework of clear, propagated artifacts rather than isolated prompts, the stateful reasoning capabilities required to succeed in these rigorous evaluations can be enhanced.

\subsubsection{The Evolution of Agentic Software Engineering}
The application of artificial intelligence in software development has shifted from simple token completion toward complex, autonomous systems. Early implementations operated primarily as advanced autocomplete engines, a paradigm that mirrors human ``System 1'' thinking: fast, instinctive, and pattern-driven \citep{kahneman2011thinking}. As Ozkaya observes, software development processes have historically improved only when automation was targeted at removing concrete roadblocks, a lesson that applies equally to generative AI tooling \citep{Ozkaya2023}. While useful for generating boilerplate code, these solutions lack structural foresight and fail when faced with complex architectural decisions. This limitation has driven a transition toward agentic software engineering, which emulates ``System 2'' thinking which is a slower, more methodical cognitive process characterized by planning, decomposition, and continuous self-correction \citep{kahneman2011thinking,Zhang2025}.
Modern architectures use multi-agent systems where dedicated AI personas (e.g., product managers, architects, testers) collaborate to solve complex problems \citep{Liu2025}. This agentic shift is reflected in multiple evaluation benchmarks like SWE-bench\footnote{\url{https://www.swebench.com/}}, which tasks models with resolving real-world GitHub issues across large codebases \citep{Jimenez2024}. SWE-bench evaluations show that pure code-generation models struggle because they lack stateful reasoning across multiple files. This demonstrates that successful automation requires a deep understanding of historical repository context and high-level design constraints. By prioritizing early intent capturing and structured documentation over simple code generation, the Sdlicit framework positions itself firmly within this next wave of reasoning-focused engineering systems.

\subsubsection{Agent Interoperability and Ecosystem Protocols}
As software systems shift from individual assistants to multi-agent networks, the mechanisms governing tool integration and communication become critical architectural choices. Traditional setups rely on ad-hoc, proprietary API integrations, which create fragile dependencies and scale poorly as new tools are introduced \citep{Ehtesham2025}. To resolve this fragmentation, emerging standards seek to formalize interoperability. One notable framework is the Agent-to-Agent (A2A) protocol, which focuses primarily on the semantic messages exchanged between distinct reasoning entities \citep{Ehtesham2025}.
While A2A helps coordinate agent dialogue, it does not explicitly address how agents interact with underlying developer infrastructure like filesystems, compilers, and database runtimes \citep{Jeong2025}.
To fill this infrastructure gap, the Sdlicit framework adopts the Model Context Protocol (MCP). MCP provides an open, client-server architecture that treats developer tools, configuration prompts, and data resources as standardized, first-class endpoints \citep{Jeong2025}. By relying on MCP, Sdlicit decouples the core LLM reasoning layer from surrounding environment tools. This allows architectural agents to query vector databases, parse local Git state, and write Gherkin test profiles using a single, uniform interface, providing a highly maintainable alternative to custom integrations \citep{Ehtesham2025}.

\subsubsection{Specification-Driven Development and Living Artifacts}
The engineering field is experiencing a rise of mentions about Specification-Driven Development (SDD), where structured documentation is treated as the primary asset of creation rather than a byproduct \citep{Babu2026}. The academic roots of this approach stem from test-driven and behavior-driven approaches taken to the extreme. In an SDD paradigm, development begins by building specifications that create strict boundaries for the automated implementation. The Sdlicit pipeline does not have this automation or SDD way of developing in mind, the goal is to create a continuous artifact lineage: Software Requirements Specifications (SRS) flow directly into Architecture Decision Records (ADRs), which can then automatically be compiled into executable BDD/Gherkin scenarios.

\subsubsection{Version Control Systems and Artifact Evolution}
Integrating Version Control Systems (VCS), specifically Git\footnote{\url{https://git-scm.com/}}, into an automated SDLC infrastructure introduces another dimension for intent propagation. A repository's commit history and pull request lineage represent a rich ledger of software engineering intent, tracking not just what code lines changed, but the rationale driving those changes via commit records and metadata. Empirical work confirms that commit message quality has a measurable impact on downstream comprehension and maintenance, and that this quality evolves over a project's lifetime \citep{Li2023}. For artifact-driven frameworks, interfacing directly with Git enables autonomous agents to parse historical iterations of design documents alongside the source files they govern. This mechanism addresses the snapshot problem directly by linking structured architectural artifacts to the project's version control baseline. Future pipelines can then analyze code commit differentials to automatically flag documentation drift, which is a recurring and well-documented failure mode in software projects \citep{Aghajani2019}. Additionally, these systems can dynamically update knowledge graphs as documentation evolves over time.

\subsubsection{Knowledge Bases for Architectural Guidance}
The effectiveness of RAG-based systems depends heavily on the quality and relevance of the underlying knowledge base.
For architectural guidance, several categories of knowledge are relevant.
First, design patterns and principles play a crucial role.
The Gang of Four patterns \citep{Gamma1994} provide reusable solutions for common software design problems, while principles such as Single Responsibility, Open/Closed, Liskov Substitution, Interface Segregation, Dependency Inversion (SOLID), Don't Repeat Yourself (DRY), Keep It Simple, Stupid (KISS), and You Aren't Gonna Need It (YAGNI) offer guidelines for maintainable code \citep{Kolhatkar2025}.
Standards and best practices provide authoritative grounding for AI suggestions.
IEEE standards such as IEEE 830 \citep{ieee830} for Software Requirements Specifications and IEEE 42010 \citep{ieee42010-2022} for architecture descriptions offer established frameworks.
Language-specific style guides and framework documentation provide context for technology-specific decisions.
This work will curate a knowledge base containing design pattern documentation, IEEE standards relevant to requirements and architecture, and technology-specific best practices.
The effectiveness of RAG-based retrieval will be evaluated against a baseline where all relevant knowledge is provided directly in the prompt, isolating the contribution of dynamic retrieval.
The system will use the knowledge base to help users create well-formed ADRs by providing relevant context from established sources.

\subsubsection{Architectural Documentation Datasets}
While numerous datasets exist for general software engineering tasks, benchmarks specifically targeting early-lifecycle architectural artifacts remain scarce. This research directly builds upon and extends the replication package produced by \citep{Buchgeher2023}. The original dataset contains 6,362 Architecture Decision Records (ADRs) extracted from 921 public GitHub repositories, providing a validated baseline for template classification and structural quality assessment. 

In this work, the \citep{Buchgeher2023} dataset serves a dual purpose. First, it functions as an empirical baseline in Phase 0 to validate template adoption metrics and continuation patterns against a broader, custom-mined landscape. Second, a curated corpus of mature ADRs (such as the \texttt{microsoft/semantic-kernel} project) is isolated from this dataset to serve as the ground-truth foundation for the system's real-world replay evaluation in Phase 2 (Section~\ref{sec:eval-sq2b}). This allows for a direct comparison between human-authored design rationale and the framework's generative output.

\newpage
\section{SDLICIT System Architecture}\label{appendix:system-architecture}

In the SDLICIT architecture, a frontend, which is either a VS~Code extension or a terminal CLI, talks to a stateless Python backend over HTTP. Short request-response calls (configuration, workspace initialisation, single-field entries) go over a REST interface. Long-running generation streams over Server-Sent Events, so the frontend can render partial output as it arrives.

The backend is a FastAPI service that also hosts an MCP server, which separates client-facing endpoints from model-facing tools. External MCP-capable agents like Claude Desktop or Cursor can therefore discover and call the same internal tools as the VS~Code extension.

\subsubsection{Architectural Decisions and Rationale}
\label{subsec:sys-decisions}

The architecture is best understood through the design decisions that produced it. Each decision in Table~\ref{tab:arch_decisions} resolves a concrete concern (testability, reproducibility, extensibility, cost) and carries a trade-off that was accepted deliberately.
The main theme is \emph{designed-in evaluability}: every subsystem is independently switchable, and this aspect is used in the evaluation in Section \ref{sec:experiments} to switch on/off different subsets of these subsystem as required by the respective experiments.

\begin{table}[!htbp]
\centering
\caption{The principal architectural decisions behind Sdlicit, the concern each resolves, and the trade-off accepted. 
}
\label{tab:arch_decisions}
\small
\scalebox{0.95}{
\renewcommand{\arraystretch}{1.5}
    \begin{tabularx}{\textwidth}{p{3.5cm} X X}
\toprule
\textbf{Decision} & \textbf{Rationale} & \textbf{Accepted trade-off} \\
\midrule
Stateless backend: the workspace directory is the only state &
The same backend can serve the CLI, the IDE extension, and external MCP agents interchangeably. It makes every run reproducible from a directory alone &
All state is written on disk and both frontend and backend need to handle it \\
 
A single orchestrator as composition root &
One assembly point with all dependencies injected, so the subsystems can be swapped or disabled for ablation &
The orchestrator is a wide class and construction order matters \\
 
A centralised LLM Gateway and prompts are declarative DSPy signatures &
Prompt changes become reviewable, diffable code. One place enforces typing, retries, model routing, and logging &
Indirection: a framework change touches every model call \\
 
Typed input/output via Pydantic, parsed in a single call (BAMLAdapter from DSPy) &
Eliminates brittle regex parsing, the schema is the single source of truth for output shape &
Schema rigidity, and malformed output triggers a re-call of the LLM (extra latency and cost) \\
 
Fragile artefacts are wrapped in self-correcting ``validated programs'' functions with DSPy &
Hard and soft quality constraints are encoded as `reward functions' that execute retries &
Up to $N$ extra model calls per artefact, the reward functions must be maintained (fine-tuning with DSPy) \\
 
A dual knowledge store sharing one retrieval graph in LightRAG &
Cross-store queries resolve in a single call, and cross-corpus graph edges are preserved &
Store isolation is a path-prefix convention, not a hard boundary \\
 
Structure-aware chunkers, one per artefact type &
Preserves the semantic boundaries (ADR sections, requirement entries) that faithful retrieval depends on &
Each new artefact type needs its own chunker \\
 
Traceability implemented as a workspace service, decoupled from the generation agents &
The one service both verifies declared trace links and infers missing ones, and computes coverage only by reading artefacts back from disk, so the system never grades its own output &
The service may parse and embed artefacts anew that the agents already produced, this is performance overhead \\
 
A three-tier Theory-of-Mind memory &
Separates cheap append-only logging from expensive LLM summarisation, enabling cross-session adaptation without unbounded context &
The upper tiers require end-of-session LLM passes \\
 
\bottomrule
\end{tabularx}
}
\end{table}

\begin{table}[t]
\centering
\caption{The major components of the Sdlicit backend and their responsibilities. Each component is either an infrastructure service or an agent. The seven agents are listed by name in Table~\ref{tab:agent_summary} and summarised individually in Section~\ref{subsec:agents}.}
\label{tab:components}
\small
\scalebox{0.85}{
\renewcommand{\arraystretch}{1.2}
\begin{tabular}{p{3.2cm} p{6.3cm} p{4.2cm}}
\toprule
\textbf{Component} & \textbf{Responsibility} & \textbf{Key collaborators} \\
\midrule
\texttt{Orchestrator} & Composition root, it owns the stage pipeline. It is the single entry point the frontend calls & Every agent and service \\
\texttt{SdlicitConfig} & Typed runtime configuration and the ablation surface. It probes the model's capabilities at start-up & \texttt{bootstrap} \\
\texttt{LLMGateway} & Routes every model call and selects the execution strategy. It manages optimised prompt state and usage accounting & DSPy and all agents \\
\texttt{KnowledgeBase} / \texttt{KBRouter} & Graph-augmented ingestion and store-aware retrieval with structure-aware chunking & LightRAG and all agents \\
\texttt{TraceService} & Traceability analysis, coverage scoring, and trace-link inference over a workspace & \texttt{LLMGateway} \\
\texttt{SocraticAgent} & Cross-cutting question-driven intervention & \texttt{LLMGateway}, ToM, KB \\
\texttt{ToMAgent} & Cross-cutting user modelling over a three-tier memory & \texttt{LLMGateway} \\
Stage agents (SOW, ADR, Requirement, User Story, BDD) & Produce the artefacts of their stage & Gateway, KB, ToM, Socratic \\
\bottomrule
\end{tabular}
}
\end{table}

\subsection{Libraries and Frameworks}
\label{subsec:sys-libraries}
 
Four external frameworks carry the architecturally significant weight. \textbf{DSPy}~\citep{Khattab2023} is the LLM-programming layer: it replaces hand-written prompt strings with typed, declarative \emph{signatures}, which is what makes a prompt change a reviewable code change and allows prompts to be optimised offline. \textbf{LightRAG}~\citep{Guo2025} is the retrieval engine. It builds a vector index and an entity-relation graph from the same corpus, giving the system both similarity search and graph-structured grounding. The \textbf{BAMLAdapter} (a DSPy component) returns schema-valid structured output in a single model call, removing a whole class of parsing code. \textbf{FastAPI} with the \textbf{Model Context Protocol} provides the backend transport and lets external coding agents drive the same pipeline. Two smaller libraries support specific guarantees: \texttt{gherkin-official}, the official Cucumber parser, validates that every generated \texttt{.feature} file is genuinely executable, and \texttt{scikit-learn} supplies the TF-IDF and topic-modelling primitives used by the semantic traceability analysis. 
 
\subsection{The Agents}
\label{subsec:agents}
 
In total there are seven separate `agents'. All share one pattern: a plain Python class with the gateway injected at construction time, async public methods, and a uniform \texttt{AgentResult} return value whose \texttt{socratic\_probe} field is populated whenever an intervention interrupts the agent's normal output. There are two agents that are cross-cutting, meaning they intervene with the others: the Socratic Solicitor and Theory of Mind. The following summaries describe what each agent does. 

\paragraph{SOW Agent (Composing).} Converts a raw client brief into a structured Statement of Work. Rather than silently inventing missing information, it routes under-specified briefs to the Socratic Solicitor and folds the resulting clarifications back in. It produces \texttt{sow.md}. The SOW is ingested into the Knowledge Base only after the engineer accepts it.
 
\paragraph{ADR Agent (Composing).} Guides the engineer through MADR-formatted Architecture Decision Records one field at a time. Each field is checked for adequacy, grounded against prior ADRs retrieved from the Knowledge Base, and may trigger a Socratic probe. A separate ``blank-page'' mode proposes candidate decisions by cross-referencing the SOW against existing ADRs to find uncovered architectural domains. It never makes the decision itself, it surfaces possible glanced-over ideas about the decision. When an ADR is saved without an explicit \texttt{implements} list, the Traceability Service infers the requirements it addresses and populates the frontmatter so trace links are established even when the engineer does not declare them by hand.
 
\paragraph{Socratic Solicitor (Expansion, cross-cutting).} Intervenes whenever an engineer's input is under-specified or a model output is ungrounded. It never proposes a solution, it asks a single question targeting a hidden assumption, a contradiction, an under-justified decision, or an unconsidered stakeholder. The goal is to keep the engineer in their Zone of Proximal Development~\citep{vygotsky1978mind} (this concept is not handled in depth, and should be part of the future work). Its judgement depth is configurable, and a second validator (the ``Dean'') refines the question when the user is detected to be struggling.
 
\paragraph{Theory of Mind (Expansion, cross-cutting).} Adapts the system's interaction style to the engineer's expertise, emotional state, and stated preferences. It maintains a three-tier on-disk memory: a raw event log, a per-session model, and an aggregated cross-session user model. Subsequently, it exposes this knowledge to the other agents as scaffolding recommendations, so probes and feedback become context-aware rather than uniform.
 
\paragraph{Requirement Agent (Composing / Generation).} Performs two jobs: a standards-grounded compliance check of an ADR against ingested ISO standards, and generation of the SRS from an approved SOW. Generated requirements are produced as a \emph{validated program}. Thus, every requirement must carry a well-formed, domain-prefixed identifier or the output is automatically re-generated.
 
\paragraph{User Story Agent (Generation).} Synthesises personas and requirements into agile user stories, each carrying first-class references to the persona it serves and the requirements it traces to. It is also a validated program: each story must carry a unique identifier and follow the canonical ``As a \ldots, I want \ldots, so that \ldots'' narrative form.
 
\paragraph{BDD Agent (Generation).} Produces two artefacts. It first generates the user \emph{personas} from the combined ADR and SRS content, then generates executable BDD scenarios as Gherkin. Gherkin generation is the most heavily validated step in the pipeline: the output is both reward-checked and parsed by the official Cucumber parser, and every \texttt{.feature} file records the requirement it verifies.
 
\medskip
\noindent Table~\ref{tab:agent_summary} consolidates the agents by stage, output, and whether their generation step is reward-validated.
 
\begin{table}[t]
\centering
\caption{The seven Sdlicit agents. 
}
\label{tab:agent_summary}
\small
\renewcommand{\arraystretch}{1.2}
\begin{tabular}{p{2.6cm} p{2.4cm} p{6.4cm} c}
\toprule
\textbf{Agent} & \textbf{Stage} & \textbf{Produces} 
\\
\midrule
SOW         & Composing   & Statement of Work (\texttt{sow.md})                 
% & --- 
\\
ADR         & Composing   & Architecture Decision Records (\texttt{adr/})       
  \\
Requirement & Comp.\,/\,Gen. & SRS (\texttt{srs.md}), ADR compliance check
  \\
User Story  & Generation  & Agile user stories (\texttt{stories.json})          
  \\
BDD         & Generation  & Personas and Gherkin scenarios (\texttt{bdd/})      
  \\
Socratic    & Expansion   & Question-driven interventions (cross-cutting)        
\\
ToM         & Expansion   & User model and scaffolding guidance (cross-cutting)  
\\
\bottomrule
\end{tabular}
\end{table}
 
\subsection{Configuration and the Ablation Surface}
\label{subsec:sys-ablation}
 
A single typed configuration object is the system's source of truth for runtime settings, and it deliberately exposes a wide \emph{ablation surface}: every architecturally significant subsystem can be switched off or swapped from one file. This is not only an operational convenience but a methodological one. This decision is what allows the evaluation in Section~\ref{sec:experiments} to isolate the contribution of each subsystem cleanly, by comparing configurations rather than maintaining separate code branches. Table~\ref{tab:config_ablation} lists the knobs and what each lets the evaluation isolate.
 
\begin{table}[t]
\centering
\caption{The configuration ablation surface. Each knob isolates one subsystem or one cost-versus-fidelity trade-off that the validation framework sweeps independently.}
\label{tab:config_ablation}
\small
\scalebox{0.75}{
\begin{tabular}{p{5cm} p{3.3cm} p{7.4cm}}
\toprule
\textbf{Knob} & \textbf{Values} & \textbf{What it isolates} \\
\midrule
\texttt{enable\_rag}        & on / off & Contribution of knowledge grounding to artefact quality. \\
\texttt{enable\_tom}        & on / off & Contribution of user-adaptive scaffolding. \\
\texttt{enable\_socratic}   & on / off & Contribution of the question-driven intervention loop. \\
\texttt{socratic\_judge\_mode} & dspy / hybrid / heuristic & Cost of LLM judgement vs.\ length-based filtering. \\
\texttt{tom\_focus}         & all / scaffolding / frustration / preferences & Per-dimension contribution of the ToM. \\
\texttt{trace\_check\_mode} & structural / semantic / full & Depth-versus-cost of traceability analysis. \\
\texttt{traceability\_graph\_source} & frontmatter / lightrag & Value of graph-extracted edges over declared ones. \\
\texttt{agentic}            & on / off & Deterministic pipeline vs.\ ReAct tool-calling agents. \\
\texttt{context\_override}  & empty / fixed text & Agent logic in isolation from the retrieval engine. \\
\bottomrule
\end{tabular}
}
\end{table}
 
\FloatBarrier

\clearpage

\section{Knowledge Retrieval and Grounding Quality}
\label{appendix:knowledge_retrieval_and_grounding_quality}

\textbf{SQ\,2} asks how architectural knowledge should be curated, structured, and retrieved to support AI-assisted decisions. To answer this, three retrieval techniques were compared on $84$ question/answer pairs derived from ISO/IEC/IEEE 29148, ISO 9241-210, and ISO/IEC/IEEE 12207.
The three techniques are:

\begin{itemize}
    \item \texttt{baseline\_direct}: no retrieval. The model answers from its own parametric memory, with no external context provided.
    \item \texttt{naive\_rag}: LightRAG's default top-$k$ chunk retrieval. The query is embedded, the closest $k$ chunks of the knowledge base are retrieved, and they are passed to the model as context. No graph traversal.
    \item \texttt{kb\_router\_hybrid}: Sdlicit's production retrieval path. It routes through LightRAG's entity-relation graph and combines graph-walked context with chunk retrieval.
\end{itemize}

\subsection{Evaluation: Four RAGAS Metrics}
\label{sec:eval-sq1-metrics}

The evaluation uses the RAGAS framework~\citep{Es2024}. RAGAS (Retrieval Augmented Generation Assessment) is a Python library that scores retrieval pipelines without needing a human-labelled answer for every question. It uses another LLM (here \texttt{openai/} \texttt{gpt-4.1-nano}) as a judge. RAGAS reports four numbers per query. Each isolates a different failure mode.

\paragraph{Faithfulness.} The judge splits the generated answer into small standalone statements and asks, for each one, whether it follows from the retrieved context. The score is the fraction of supported claims. A faithfulness of $1.0$ means every claim is grounded. A faithfulness of $0.0$ means none are. This is the metric on which retrieval should show the biggest effect. If it does not, the retrieved context is decoration rather than grounding.

\paragraph{Answer Relevancy.} The judge generates three synthetic questions from the answer, embeds them with \texttt{text-embedding-3-large}, and takes their mean cosine similarity to the original user question. A score near $1$ means the answer is on topic. The catch is that LLMs are fluent and on topic by default, so a high answer relevancy does not imply correctness. Both grounded and ungrounded answers can score well here.

\paragraph{Context Recall.} The judge splits the gold reference answer, not the model's generated answer, into atomic claims and asks how many of those claims are recoverable from the retrieved context. This measures how much of the right information made it into the prompt. It is the retrieval side metric.

\paragraph{Context Precision.} For each retrieved chunk, the judge asks whether the chunk is relevant to the question. The score rewards relevant chunks appearing early in the retrieval ranking. 

\begin{table}[!htbp]
\centering
\caption{Aggregated RAGAS scores, mean and standard deviation, over $N=84$ ISO derived gold questions. 
}
\label{tab:ragas_summary}
\resizebox{\linewidth}{!}{%
\begin{tabular}{lcccc}
\toprule
\textbf{Technique} 
& \textbf{Faithfulness} 
& \textbf{Answer Rel.} 
& \textbf{Context Recall} 
& \textbf{Context Prec.} \\

\midrule
\texttt{baseline\_direct} (no retrieval) 
& $0.007$ \,($0.056$) 
& $0.780$ \,($0.191$) 
& $0.133$ \,($0.333$) 
& $0.024$ \,($0.153$) \\

\texttt{kb\_router\_hybrid}              
& $\mathbf{0.974}$ \,($0.053$) 
& $0.754$ \,($0.160$) 
& $\mathbf{0.744}$ \,($0.307$) 
& $0.773$ \,($0.277$) \\

\texttt{naive\_rag}                      
& $\mathbf{0.979}$ \,($0.052$) 
& $0.731$ \,($0.215$) 
& $0.666$ \,($0.329$) 
& $\mathbf{0.817}$ \,($0.333$) \\
\bottomrule
\end{tabular}%
}
\end{table}

\subsection{Results and analysis} Looking at the results in Table \ref{tab:ragas_summary}, note that the \texttt{baseline\_direct} configuration bypasses retrieval entirely and answers from parametric weights alone. It yields a faithfulness score of $0.007$, essentially zero, while both retrieval-augmented techniques jump above $0.97$. This is the largest single effect in the entire validation. It is also the strongest evidence that retrieval is doing real work, rather than dressing the answer with citations after the fact. The standard deviations across the $84$ questions are tight (about $0.05$), so the improvement is uniform across the corpus and not driven by a few lucky generations.

\paragraph{Limitations }A limitation of the RAG-based evaluation is that it looks at the performance of the LLM in combination with the (LightRAG) retrieval strategy, but performs the evaluation using a generated set of queries and answer pairs, using the knowledge-base documents as the input for the question-answer generation. While this does provide a domain-specific test of retrieval performance with respect to LLM generation, it cannot not evaluate the effect on retrieval and generation within the full system. 

\clearpage

\section{Primary Scenario: Time-Tracking Application}
\label{sec:appendix-timetrack-brief}

\scalebox{0.9}{
\begin{custombox}[title=Client Interview Summary -- Time-Tracking Application, width=\linewidth]
\small
Hi, so I'm from the research group, and we want to talk about building something for time tracking. Let me give you context first because it'll matter.
We mostly run funded projects for companies. These are usually feasibility studies or proofs of concept the kind of work a company does before it commits to a real investment. We're not profit-driven. The goal is to stay afloat and keep the lab running. Nobody's gaming margin here; we're just trying to keep researchers employed and on a job.
For users: we'd start with my own group, which is around 7 people some full-time, some not. There are two other research groups in the same facility, around 10 more people total, who'd also try it. If that goes well, I'd take it to management and propose rolling it out across all research groups company-wide, which is somewhere between 40 and 60 people. But that's a long way out don't design for it now.
The actual problem. We currently use a basic timesheet tool I'd rather not name it, but we're used to it. The issue is that for subsidized projects we're allocated a fixed number of billable hours per researcher per project. Say a project gives me 400 hours. In a consultancy you'd just bill what you actually work, but our setup is different. Researchers carry a lot of overhead meetings, internal coordination, other projects and so the relationship between time-actually-spent and time-allocated is loose. If a researcher works 150 hours on something budgeted at 200, they end up filling in 200 anyway, because that's what we have to report to the funder. The current timesheet doesn't reflect reality, and the researchers know it. They hate filling it in, and then admin still comes back and asks them for more hours because they didn't hit the target. It's a double tax on their time. We also occasionally have to export to Excel in formats specific to a particular funder, so whatever we build has to be adaptable on the output side.
What we want: a system that helps researchers fill in their billable hours sensibly. When someone is supposed to log 100 hours on a topic and they've only naturally logged 80, the system should help them close the gap using context from what they actually did meetings from Outlook, commits and pushes from GitHub, possibly activity in other tools like Word. The point isn't surveillance. We're not trying to monitor what people are working on. The output is for the developer themselves it's a draft they review before submitting. The thing we want to kill is the back-and-forth with admin asking for more hours after the timesheet is already in.
What we don't need: a production system. This is a pilot. No high availability, no multi-tenancy, no 24/7 uptime. We need something the small group can run for real for a few months and generate data from, that we can build on later.
A lot of things are open. Hosting undecided. Output format depends on the funder. Whether the team lead or research coordinator should see anyone else's data currently leaning toward no, only the individual sees their own draft, but that might change. For GitHub we're not sure what to actually look at: diffs, just commits, push timing, end-of-project repo sweep each has weaknesses, especially if someone doesn't push frequently. We need some way to link a project to its repos and to its Outlook events automatically, probably via keywords or some LLM-based classification, but the exact mechanism is open. We'll need a place to feed in project context keywords, sub-case names, anything that helps disambiguate when projects overlap.
Done enough to use: a frontend where I enter project info, a backend that ingests from Outlook and GitHub and uses an LLM to propose time attributions, and a view of my hours by project I can review and edit. That's the bar.
\end{custombox}
}

\clearpage
\section{Multi-model Experiments}
\label{appendix:multimodel_experiments}

\begin{table}[!htbp]
\centering
\caption{Per-model results from the multi-model evaluation, $N=5$ runs per model. 
``Err.'' is the mean count of pipeline-stage errors per run.
}
\label{tab:model_comparison}
\resizebox{\linewidth}{!}{%
\begin{tabular}{llcccccr}
\toprule
\multirow{4}{*}{\textbf{Model}} & \multirow{4}{*}{\textbf{Backend}} & \multirow{4}{2cm}{\textbf{Struct. trace coverage (\%)}} & \multirow{4}{2cm}{\textbf{Sem. trace coverage (\%)}} &  \multirow{4}{1.5cm}{\textbf{Gherkin validity (\%)}} & \multirow{4}{*}{\textbf{Tokens}} & \multirow{4}{*}{\textbf{Latency (s)}} & \multirow{4}{*}{\textbf{Err.}} \\
 & & & & & & & \\
 & & & & & & & \\
 & & & & & & & \\
 & & $\mu$ ($s$) & $\mu$ ($s$) & $\mu$ ($s$) & $\mu$ & $\mu$ ($s$) & $\mu$ \\
 
\midrule
gpt-5.4-mini       & OpenRouter (cloud)   & $100.0$ ($0.0$)  & $45.0$ ($4.9$)   & $100.0$ ($0.0$)  & ${\sim}33{,}200$ & $806$ ($56$)    & $0.0$ \\
gpt-4o             & OpenRouter (cloud)   & $100.0$ ($0.0$)  & $47.0$ ($19.9$)  & $100.0$ ($0.0$)  & ${\sim}27{,}300$ & $527$ ($29$)    & $0.0$ \\
mistral-small      & OpenRouter (cloud)   & $100.0$ ($0.0$)  & $41.0$ ($37.9$)  & $60.0$ ($54.8$)  & ${\sim}28{,}400$ & $733$ ($205$)   & $0.4$ \\
qwen3.5-4b-sdlicit & Ollama (local 4B)    & $86.3$ ($14.2$)  & $30.7$ ($31.0$)  & $93.3$ ($14.9$)  & ${\sim}107{,}700$ & $2{,}294$ ($295$) & $1.4$ \\
qwen3.5-0.8b-sdlicit & Ollama (local 0.8B)& $0.0$ ($0.0$)    & $0.0$ ($0.0$)    & $0.0$ ($0.0$)    & ${\sim}91{,}500$ & $694$ ($257$)   & $2.6$ \\
lfm2.5-1.2b-sdlicit & Ollama (local 1.2B) & $46.7$ ($7.5$)   & $0.0$ ($0.0$)    & $0.0$ ($0.0$)    & ${\sim}20{,}800$ & $299$ ($25$)    & $4.0$ \\
\bottomrule
\end{tabular}%
}
\end{table}

\clearpage
\section{Fuzzy Jaccard on the External Replay: A Worked Example}
\label{app:jaccard-worked}

\begin{table}[h]
\centering
\caption{Pairwise character-level similarity (Ratcliff--Obershelp ratio) between the two ground-truth options ($A_1, A_2$) and the three generated alternatives ($B_1$--$B_3$) of ADR-0006. The maximum, $0.273$, is highlighted. It is still well below the $\tau = 0.6$ matching threshold, so the fuzzy Jaccard collapses to zero.}
\label{tab:jaccard_matrix}
\small
\begin{tabular}{lccc}
\toprule
 & $B_1$ & $B_2$ & $B_3$ \\
\midrule
$A_1$ & $0.050$ & $0.216$ & $0.179$ \\
$A_2$ & $0.055$ & $\mathbf{0.273}$ & $0.220$ \\
\bottomrule
\end{tabular}
\end{table}

The Semantic Kernel replay (Section~\ref{sec:eval-sq2b}, Table~\ref{tab:sk_jaccard}) reports a fuzzy Jaccard of $0.0$ on the alternatives sets of all ten ADRs. This appendix unpacks why, on one representative record, so that the zero is read as a property of the metric and the data rather than a generator failure.

\paragraph{What is compared.} The fuzzy Jaccard is computed over two sets of \emph{whole option strings}: $A$, the ground-truth \texttt{considered\_options} of the real Semantic Kernel ADR, and $B$, the \texttt{\#\# Alternatives} bullets Sdlicit generated. There is no tokenisation and no stop-word filtering. Each option string is compared to each other option string character by character with \texttt{difflib.SequenceMatcher.ratio} (the Ratcliff--Obershelp ratio), case-folded. A pair counts as a match when one string contains the other or the ratio reaches $\tau = 0.6$. This is a different measure on a different artefact from the stop-word-filtered keyword Jaccard used inside the semantic trace-coverage composite (Section~\ref{sec:eval-sq2a-metrics}). The two should not be conflated.

\paragraph{ADR-0006 (Dynamic REST payload building).} The real ADR considered two options, and Sdlicit generated three:

\begin{itemize}
  \item[] \textbf{Ground truth $A$:}
  \begin{itemize}
    \item[$A_1$] ``Enable the dynamic creation of payload and/or namespacing by default.''
    \item[$A_2$] ``Enable the dynamic creation of payload and/or namespacing based on configuration.'' \quad($\star$ chosen)
  \end{itemize}
  \item[] \textbf{Generated $B$:}
  \begin{itemize}
    \item[$B_1$] ``Free-form payload assembly: directly merge arbitrary JSON objects from workflow/model/tool outputs into a final payload without schema-driven field mapping or namespacing, collisions resolved by last-write-wins.''
    \item[$B_2$] ``Global flat parameter namespace: require all clients to manually prefix all keys (e.g.\ \texttt{tool\_}, \texttt{workflow\_}) and rely on string-key conventions rather than endpoint-contract-driven mapping.''
    \item[$B_3$] ``No deterministic mapping layer: pass raw model/tool arguments to REST calls and only validate server responses, rely on server-side validation and errors for correction.''
  \end{itemize}
\end{itemize}

The full $2\times3$ matrix of pairwise Ratcliff--Obershelp ratios is given in Table~\ref{tab:jaccard_matrix}. The largest entry anywhere in the matrix is $0.273$ ($A_2$ vs.\ $B_2$), which is the \texttt{Max sim} value reported for ADR-0006 in Table~\ref{tab:sk_jaccard}. Because $0.273 < 0.60$, not a single pair clears the threshold, the intersection is empty, and the fuzzy Jaccard is exactly $0$.

\paragraph{Why this is a metric artefact, not a reasoning failure.} Two structural effects guarantee the zero independently of phrasing. First, the two sides describe the same concepts in different words and at different lengths, so even the closest pair shares too little verbatim character overlap to clear $\tau$. Second, the sets are not even drawn from the same population: the ground-truth list retains the \emph{chosen} option ($A_2$), whereas Sdlicit's \texttt{\#\# Alternatives} section lists only the \emph{rejected} options, so the one option both sides agree on can never appear on both sides to be matched. This is exactly why the decision-level LLM judge, which reads the chosen decision rather than the rejected-option vocabulary, rules ADR-0006 a \texttt{match} even though its fuzzy Jaccard is $0$. The fuzzy Jaccard is therefore reported strictly as a surface-vocabulary indicator, and the decision-agreement label carries the substantive SQ\,3 replay claim.

\clearpage
\section{Semantic Kernel Replay: Real versus Generated Architecture Decisions}
\label{app:sk-decisions}

The worked example above (Appendix~\ref{app:jaccard-worked}) explains \emph{why} the fuzzy Jaccard collapses to zero on every Semantic Kernel record: the metric reads rejected-option vocabulary, not the chosen decision. This appendix completes the picture from the other side. It places the real human ADR next to the generated one for two records, so the reader can read the actual artefacts and see what the decision-agreement judge is responding to. Both records score a fuzzy Jaccard of $0.00$ (as do all ten, Table~\ref{tab:sk_jaccard}), yet at the decision level one is a clean \texttt{match} and the other is a \texttt{different-justifiable} divergence. The contrast shows that the substantive SQ\,3 claim lives in the decision agreement, not in the surface metric. All text below is verbatim from the replay artefacts in \texttt{notebooks/output/nb08} and \texttt{notebooks/intermediate/nb08/sk\_unified}.

\subsection{ADR-0017: OpenAI Function Calling (decision \texttt{matches}, Jaccard $0.00$)}

Here the human author and Sdlicit reach the same decision in different words. The closest pairwise option similarity is $0.42$, the highest in the whole study (Table~\ref{tab:sk_jaccard}), but it is still below the $\tau = 0.6$ threshold, so the thresholded fuzzy Jaccard is $0.00$.

\begin{itemize}
  \item[] \textbf{Real ADR (\texttt{microsoft/semantic-kernel}).}
  \begin{itemize}
    \item \emph{Considered options.} (i)~support sending/receiving functions via the chat-completions endpoint \emph{with} modifications to the interfaces, (ii)~support it \emph{without} modifications to the interfaces, (iii)~implement a planner around the function-calling capability.
    \item \emph{Decision.} Support sending/receiving functions via the chat-completions endpoint \emph{without} modifications to the interfaces. ($\star$ chosen)
  \end{itemize}
  \item[] \textbf{Generated ADR (Sdlicit).}
  \begin{itemize}
    \item \emph{Alternatives.} Hard-code OpenAI-specific function-calling semantics into the orchestration core; implement it only as a client-side integration without platform governance hooks; rely solely on plain-text (regex) tool-call extraction; support function calling only for a single-tool workflow.
    \item \emph{Decision.} Implement OpenAI function calling as a provider adapter that conforms to the platform's common request/response abstractions, confining OpenAI-specific payload mapping to the OpenAI provider layer and keeping the orchestration core model-agnostic. ($\star$ chosen)
  \end{itemize}
\end{itemize}

The two chosen decisions are the same architectural move: add function-calling support \emph{without} disturbing the shared interfaces, by isolating the provider-specific translation. The judge labelled this record a \texttt{match}, with the rationale that ``the generated decision is effectively the same as the human choice: implement OpenAI function calling through a provider adapter that fits the existing common abstractions, which corresponds to supporting it without modifying interfaces.'' The rejected-option lists, however, share no vocabulary: the human options are phrased as interface-modification trade-offs, while the generated alternatives enumerate concrete anti-patterns. That is why the alternatives quality is \texttt{divergent} and the fuzzy Jaccard is $0.00$ even though the decision is a match.

\subsection{ADR-0051: Entity Framework Vector Store (decision \texttt{different-justifiable}, Jaccard $0.00$)}

Here the two sides reach \emph{opposite} decisions, yet both are internally coherent. The closest pairwise option similarity is $0.35$ and the fuzzy Jaccard is again $0.00$.

\begin{itemize}
  \item[] \textbf{Real ADR (\texttt{microsoft/semantic-kernel}).}
  \begin{itemize}
    \item \emph{Considered options.} (i)~add a new \\ \texttt{Microsoft.SemanticKernel.Connectors.EntityFramework} connector, (ii)~do not add it, and instead add a connector for an individual database only when needed.
    \item \emph{Decision.} Do \emph{not} add the Entity Framework connector; add database-specific connectors as the need arises. ($\star$ chosen)
  \end{itemize}
  \item[] \textbf{Generated ADR (Sdlicit).}
  \begin{itemize}
    \item \emph{Alternatives.} Use a dedicated vector database (pgvector, Pinecone, Weaviate, Qdrant) via its native SDK; store vectors outside the relational model and retrieve via an external index, using EF only for metadata; use raw ADO.NET/SQL without EF; use in-memory indexing with periodic snapshots.
    \item \emph{Decision.} Implement an EF-based vector-store connector as a pluggable memory/knowledge provider, packaged as a versioned connector with governance, secret handling, audit logging, and OpenTelemetry instrumentation. ($\star$ chosen)
  \end{itemize}
\end{itemize}

This is the opposite decision: the human declined the EF connector while Sdlicit chose to build one. The judge labelled the record \texttt{different-justifiable}, noting that ``the generated decision chose the opposite of the human decision\ldots however, the generated choice is still technically coherent and reasonable, with a plausible implementation approach.'' The option sets overlap only partially (\texttt{alternatives quality: partial}), and the fuzzy Jaccard is $0.00$.

\subsection{What the Two Records Show Together}

The two records bracket the SQ\,3 result. ADR-0017 is a decision \texttt{match}; ADR-0051 is a defensible divergence. Their fuzzy Jaccard scores are identical at $0.00$, and so is the score of every other Semantic Kernel record. The metric therefore cannot distinguish a clean agreement from an opposite-but-coherent choice, because it only reads rejected-option vocabulary, and on architectural-pattern prose that vocabulary never aligns. The decision-agreement judge \emph{can} make that distinction, which is exactly why the chapter reports the fuzzy Jaccard strictly as a surface-vocabulary indicator and lets the decision-agreement label carry the substantive replay claim.

\subsection{The Generated SRS}

The replay also produces a Software Requirements Specification from the same repository context, stored at \\ \texttt{notebooks/intermediate/nb08/sk\_unified/.sdlicit/artifacts/srs.md}. \\ The Semantic Kernel repository ships no formal requirements specification, so the generated SRS has no human counterpart to compare against. It is included here to show the shape of the generated artefact: each requirement carries a stable identifier that the downstream ADRs cite in their \texttt{implements} front-matter (for instance the function-calling ADR above implements \texttt{REQ-PROTOCOL-04}). This is the structure the Traceability Service later reads back when it resolves trace links.

\begin{quote}\small
\textbf{[REQ-MODEL-03]} The platform shall provide a model-agnostic provider interface that allows developers to configure and use multiple AI providers using a common request/response abstraction.

\textbf{[REQ-PROTOCOL-04]} The platform shall support tool/function calling by allowing developers to register callable tools with typed input/output contracts and by executing tool calls requested by the model.

\textbf{[REQ-OUTPUT-05]} The platform shall provide first-class structured outputs based on developer-declared schemas (e.g.\ JSON Schema or typed contracts) and shall validate outputs before returning them to the caller.

\textbf{[REQ-MEMORY-08]} The platform shall provide pluggable memory/knowledge grounding via configurable providers that support retrieval of relevant prior context or enterprise content.
\end{quote}

\end{subappendices}

\end{document}